\documentclass[preprintnumbers,eqsecnum,aps,prd,epsf,showpacs,nofootinbib]{revtex4}
\usepackage{latexsym,amsmath,amssymb}
\usepackage[dvips]%[dvipdfm]
{graphicx,color}% Include figure files
\usepackage{dcolumn,bm}    % Align table columns on decimal point
\usepackage{subfigure}  % sub labeling of figures
\newcommand{\gsim}{\mbox{\raisebox{-1.ex}{$\stackrel
      {\textstyle>}{\textstyle\sim}$}}}

\def\til#1{\tilde{#1}} 
\definecolor{red}{rgb}{1,0,0}
\definecolor{blue}{rgb}{0,0,1}
\definecolor{green}{rgb}{0,1,0}
\begin{document}
\thispagestyle{empty}
\title{Non-Gaussianity of superhorizon curvature perturbations 
beyond $\delta$ N formalism}

\preprint{RESCEU-6-10}
\preprint{IPMU-10-0042}
\preprint{YITP-10-8}

\author{Yu-ichi Takamizu$^{1}$}
\email{takamizu_at_resceu.s.u-tokyo.ac.jp}

\author{Shinji Mukohyama$^{2}$}
\email{shinji.mukohyama_at_ipmu.jp}

\author{Misao Sasaki$^{3}$}
\email{misao_at_yukawa.kyoto-u.ac.jp}

\author{Yoshiharu Tanaka$^{3}$}

\affiliation{
\\
$^{1}$ Research Center for the Early Universe (RESCEU), Graduate
School of Science, The University of Tokyo, Tokyo 113-0033, Japan 
\\
$^{2}$ Institute for the Physics and Mathematics of the Universe (IPMU),
The University of Tokyo, 5-1-5 Kashiwanoha, Kashiwa, 
Chiba 277-8582, Japan
\\
$^{3}$ Yukawa Institute for Theoretical Physics
      Kyoto University, Kyoto 606-8502, Japan}
\date{\today}

\begin{abstract}
We develop a theory of nonlinear cosmological perturbations on
superhorizon scales for a single scalar field with a general kinetic term 
and a general form of the potential. We employ the ADM formalism and the spatial
 gradient expansion approach, characterised by $O(\epsilon^m)$, 
where $\epsilon=1/(HL)$ is a small parameter representing the ratio of the 
Hubble radius to the characteristic length scale $L$ of perturbations. 
We obtain the general solution for a full nonlinear version
of the curvature perturbation valid up through second-order in $\epsilon$ ($m=2$). 
We find the solution satisfies a nonlinear second-order differential
equation as an extension of the equation for the linear curvature
perturbation on the comoving hypersurface.
Then we formulate a general method to match a perturbative solution 
accurate to $n$-th-order in perturbation inside the horizon 
to our nonlinear solution accurate to second-order ($m=2$) in
the gradient expansion on scales slightly greater than the Hubble radius.
The formalism developed in this paper allows us to calculate the superhorizon 
evolution of a primordial non-Gaussianity beyond the so-called $\delta N$
formalism or separate universe approach which is equivalent
to leading order ($m=0$) in the gradient expansion.
In particular, it can deal with the case when there is
a temporary violation of slow-roll conditions.
As an application of our formalism, 
we consider Starobinsky's model, which is a single field model 
having a temporary non-slow-roll stage due to a sharp change in
the potential slope. 
We find that a large non-Gaussianity can be generated even on superhorizon
scales due to this temporary suspension of slow-roll inflation.
\end{abstract}
\maketitle

\section{Introduction}
\label{sec:intro}

Recent observations of the cosmic microwave background anisotropy
show very good agreement of the observational data with
the predictions of conventional, single-field slow-roll models of
inflation, that is, adiabatic Gaussian random primordial fluctuations with
an almost scale-invariant spectrum~\cite{Spergel:2006hy,Komatsu:2010fb}.
Nevertheless, as the observational accuracy improves,
it has become observationally feasible to detect a small non-Gaussianity
in the data~\cite{Bartolo:2004if,Komatsu:2001rj,Komatsu:2010fb}.
In particular, the PLANCK satellite \cite{Planck:2006uk} launched last year
is expected to bring us much finer data and it is hoped that
non-Gaussianity may actually be detected.
As a consequence, non-Gaussianity from inflation has been a focus of much
attention in recent 
years~\cite{Seery:2005wm,Seery:2005gb, Rigopoulos:2003ak,Lyth:2005fi}. 

To study possible origins of non-Gaussianity, one must go beyond
the linear perturbation theory~\cite{Sasaki:1998ug,Lyth:2004gb,Langlois:2006vv}.
The conventional models of inflation cannot explain an observationally
detectable level of non-Gaussianity, since the magnitude of it
is extremely small, suppressed by slow-roll parameters~\cite{Maldacena:2002vr}.
Then a variety of ways to generate a large non-Gaussianity 
have been proposed. They may be roughly classified into two;
multi-field models that produce non-Gaussianity classically on superhorizon 
scales~\cite{Suyama:2007bg,Sasaki:2008uc,Malik:2006pm,Sasaki:2006kq,Yokoyama:2007uu, Byrnes:2008wi},
and non-canonical kinetic term models that produce non-Gaussianity quantum
mechanically on subhorizon
scales~\cite{Alishahiha:2004eh,Chen:2006nt,Chen:2006xjb}.
In particular, in the former case, 
the $\delta N$ formalism~\cite{Starobinsky:1986fxa,Sasaki:1995aw,Wands:2000dp}
turned out to be a powerful tool for the estimation of 
non-Gaussianity~\cite{Sasaki:1998ug,Rigopoulos:2003ak,Lyth:2004gb}.

In order to parameterize the amount of non-Gaussianity of 
primordial perturbations, the nonlinear parameter $f_{NL}$ 
is commonly used \cite{Komatsu:2001rj,Spergel:2006hy}.
 This is related to the bispectrum of the curvature perturbation 
on $\zeta$~\cite{Wands:2000dp},
 and is generally defined as
\begin{eqnarray}
f_{NL}={5\over 6}{\prod_{i=1}^3 k_i^3 \over \sum_{i=1}^3 k_i^3}
{B_{\zeta}({\bm k}_1,{\bm k}_2,{\bm k}_3)\over 4\pi^4 {\cal P}_{\zeta}^2 }\,.
\end{eqnarray}
Here ${\cal P}_\zeta$ and $B_{\zeta}$ are 
the power spectrum and bispectrum of $\zeta$, respectively, 
and they are defined in Fourier space by 
\begin{eqnarray}
\langle {\zeta}_{{\bm k}_1} {\zeta}_{{\bm k}_2} \rangle&=&
(2\pi)^3 \delta^3({\bm k}_1+{\bm k}_2){2\pi^2 \over k_1^3} 
{\cal P}_{\zeta}({\bm k}_1)\,, \nonumber\\
\langle {\zeta}_{{\bm k}_1} {\zeta}_{{\bm k}_2}{\zeta}_{{\bm k}_3} 
\rangle&=&(2\pi)^3 \delta^3({\bm k}_1+{\bm k}_2+{\bm k}_3)
B_{\zeta}({\bm k}_1,{\bm k}_2,{\bm k}_3)\,,
\end{eqnarray}
Corresponding to the two different origins of non-Gaussianity mentioned
above, the nonlinear parameter $f_{NL}$ can be mainly 
classified into two types; the {\it local\/} type, $f_{NL}^{\rm local}$,
which may arise from multi-scalar models on superhorizon scales,
and {\it equilateral\/} type, $f_{NL}^{\rm equil}$, which
arises from non-canonical kinetic term models on subhorizon
 scales.\footnote{A new type of $f_{NL}$ has been studied 
recently~\cite{Senatore:2009gt}, called the {\it orthogonal\/}
type. This may be generated from higher derivative terms
in the action.} 
%

%%%%%%%%%%%%%%%%%%%
The local type is called so because it represents
a local, point-wise non-Gaussianity given by
\begin{eqnarray}
{\zeta}({\bm x})={\zeta}_G({\bm x})+{3\over 5}f_{NL}^{\rm local} 
{\zeta}^2_G({\bm x})\,,
\end{eqnarray}
where ${\zeta}_G$ is the Gaussian random field.
On the other hand, the equilateral form of 
the bispectrum is given by
\begin{eqnarray}
f_{NL}^{\rm equil}({\bm k}_1, {\bm k}_2, {\bm k}_3)
={10\over 3} {{\cal A}_{NL}\over \sum_i k_i^3},
\end{eqnarray}
with the shape function ${\cal A}_{NL}$ typically in the 
form~\cite{Chen:2006nt},
\begin{eqnarray}
{\cal A}_{NL}\propto {1\over 8}\sum_i k_i^3 -{1\over K}\sum_{i<j} k_i^2 
k_j^2+{1\over 2K^2} \sum_{i\neq j} k_i^2 k_j^3,~~~~{\rm with\ \ } 
K\equiv k_1+k_2+k_3.
\end{eqnarray}
Note that the sign convention of $f_{NL}$ here follows WMAP's sign convention 
and it is opposite to Maldacena~\cite{Maldacena:2002vr}. 
On the observational side, the current bounds on the parameter $f_{NL}$ by 
WMAP seven years~\cite{Komatsu:2010fb} are 
$-10<f_{NL}^{\rm local}<74$ (95\% C.L.) for the local
form of the bispectrum and $ -214<f_{NL}^{\rm equil}<266$ (95\% C.L.)
for the equilateral form. 
By PLANCK~\cite{Planck:2006uk}, 
it is expected that non-Gaussianity of the level $|f_{NL}|\gsim 5$ can be
detected~\cite{Komatsu:2001rj}. 

In this paper, we investigate another possible origin of non-Gaussianity,
namely, non-Gaussianity due to a temporary non-slow roll stage
on superhorizon scales. In order to investigate such a case,
however, the $\delta N$ formalism is not sufficient. The reason
is as follows. On superhorizon scales, one can employ
the spatial gradient
 expansion~\cite{Starobinsky:1986fxa,Nambu:1994hu,Salopek:1990jq, Kodama:1997qw},
which is characterised by the expansion parameter $\epsilon=1/(HL)$ representing
the ratio of the Hubble horizon radius to the characteristic 
length scale $L$ of the perturbation. 
The $\delta N$ formalism or the separate universe approach 
is equivalent to the leading order approximation, i.e., $O(\epsilon^0)$
in the gradient expansion. It is valid at a slow-roll stage when
local values of the inflaton field at each local point
(averaged over each horizon-size region) determine the evolution of
the universe at each point. In the context of perturbation theory,
this implies one can ignore the decaying mode of the curvature perturbation.
However, when the slow-roll condition is violated, 
the decaying mode cannot be neglected any more and the gradient expansion
to $O(\epsilon^2)$ is known to play a crucial role already at the level
of linear perturbation theory~\cite{Seto:1999jc,Leach:2001zf,Jain:2007au}.
Thus, to evaluate non-Gaussianity from a non-slow-roll stage
of inflation, it is necessary to develop a nonlinear theory of
cosmological perturbations valid up through $O(\epsilon^2)$ in the gradient
expansion~\cite{Tanaka:2006zp,Tanaka:2007gh,Takamizu:2008ra}. 

This paper is organized as follows. In Sec.~\ref{sec:gradexp}, 
we briefly review the spatial gradient expansion to $O(\epsilon^2)$ in the
uniform Hubble slicing~\cite{Tanaka:2006zp,Tanaka:2007gh,Takamizu:2008ra}. 
In Sec.~\ref{sec:nlcurvpert}, we develop a theory of full nonlinear 
curvature perturbations on superhorizon scales valid up through
 $O(\epsilon^2)$ in the gradient expansion.
In doing so, we introduce a variable that represents the nonlinear 
curvature perturbation as an extension of the linear comoving curvature
perturbation. Then we derive an explicit expression for the
nonlinear solution. We find a nonlinear
second-order differential equation which the general solution satisfies.
In Sec.~\ref{sec:matching}, we develop a general formalism 
for matching a $n$-th-order perturbative solution 
to the general nonlinear solution we found on superhorizon scales.
Then in Sec.~\ref{sec:linearmatch} we consider a special case
when a linear perturbative solution is matched to
the nonlinear solution. This applies to the case when the inflaton is 
still slow-rolling when the scale of interest crosses the Hubble horizon
scale. As an application of our formalism, we study Starobinsky's model
in Sec.~\ref{sec:starobinsky}.  
Sec.~\ref{sec:summary} is devoted to summary and discussion. 

\section{Gradient expansion in uniform Hubble slicing}
\label{sec:gradexp}
In this section, we briefly review theory of 
nonlinear cosmological perturbations valid up through $O(\epsilon^2)$ 
in the spatial gradient expansion, following~\cite{Takamizu:2008ra}.
Throughout this paper we consider Einstein gravity and
a minimally-coupled single scalar field described by the action,
%============< EQUATION >==============%
%
\begin{equation}
 I = \int d^4x\sqrt{-g}\left[\frac{R}{16 \pi G_N}+P(X,\phi)\right]\,, 
\end{equation}
%======================================%
where $R$ is the Ricci scalar and 
$X=-g^{\mu\nu}\partial_{\mu}\phi\partial_{\nu}\phi$. The reason
why the scalar field Lagrangian is denoted by $P$ is that
it plays the role of the pressure as shown immediately below.
We assume $\partial_\mu\phi$ is timelike. Then the stress energy 
tensor of the scalar field may be put in the perfect fluid form,
%============< EQUATION >==============%
%
\begin{equation}
 T_{\mu\nu} = 2P_X\partial_{\mu}\phi\partial_{\nu}\phi + Pg_{\mu\nu}
  = (\rho+P)u_{\mu}u_{\nu} + Pg_{\mu\nu}\,,
  \label{eqn:Tmunu-phi}
\end{equation}
%======================================%
where the energy density $\rho$ and the 4-velocity $u^\mu$ are 
given by
%============< EQUATION >==============%
%
\begin{equation}
 \rho(X,\phi) = 2P_X X - P, \quad
  u_{\mu} = -\frac{\partial_{\mu}\phi}{\sqrt{X}}\,.
 \end{equation}
Hereafter, 
the subscripts $X$ and $\phi$ represent derivative with respect to
$X$ and $\phi$, respectively. 
%======================================%
The following relation among the first-order variations of $P$, $\rho$ and
$\phi$ will be useful in the analysis below:
%============< EQUATION >==============%
%
\begin{equation}
 \delta P  = c_s^2\delta\rho + \rho\Gamma\delta\phi\,,
  \label{eqn:deltaP}
 \end{equation}
%======================================%
where
%============< EQUATION >==============%
%
\begin{equation}
 c_s^2 = \frac{P_X}{2P_{XX}X+P_X}\,, \quad
  \Gamma = \frac{1}{\rho}\left(P_{\phi}-c_s^2\rho_{\phi}\right). 
\end{equation}
%======================================%
Note that $c_s$ is the speed of sound for a gauge-invariant scalar
perturbation in linear theory~\cite{Garriga:1999vw}. 

We consider a theory of nonlinear cosmological perturbations on
superhorizon scales developed in~\cite{Tanaka:2007gh,Takamizu:2008ra}. 
We employ the ADM formalism and the spatial gradient expansion in 
the {\it uniform Hubble slicing}. To make clear the relation between
the standard perturbative expansion and the spatial gradient expansion,
let us introduce the two numbers $n$ and $m$
associated with the two different expansions: 
The number $n$ denotes the order in the standard perturbative expansion
with respect to the amplitude of perturbation,
and the number $m$ denotes the order in the spatial gradient
expansion, $O(\epsilon^m)$.
For example, linear cosmological perturbation theory corresponds
to ($n=1, m=\infty$), and second-order perturbation theory to 
($n=2, m=\infty$). The $\delta N$ formalism corresponds to ($n=\infty, m=0$).
Our study corresponds to ($n=\infty, m=2$).

In the ADM decomposition, the metric is expressed as 
\begin{equation}
 ds^2 = - \alpha^2 dt^2 + \gamma_{ij}(dx^i+\beta^idt)(dx^j+\beta^jdt), 
\end{equation}
where $\alpha$ is the lapse function, $\beta^i$ is the shift vector and
the Latin indices run over $1, 2, 3$. We also need the extrinsic
curvature $K_{ij}$ defined by 
%============< EQUATION >==============%
%
\begin{equation}
 K_{ij} =
  -\frac{1}{2\alpha}\left(\partial_t\gamma_{ij}-D_i\beta_j-D_j\beta_i\right)\,,
  \label{eqn:def-K}
\end{equation}
where $D$ is the covariant derivative with the spatial metric $\gamma_{ij}$.
Then the variations of the action with respect to $\alpha$ and $\beta^i$ 
lead to constraint equations, namely, the Hamiltonian and momentum constraints,
respectively, while the variation with respect to 
the spatial metric $\gamma_{ij}$ gives dynamical equations,
which may be written as a set of first-order differential
equations for $\gamma_{ij}$  and $K_{ij}$.
For convenience, we further 
decompose the spatial metric and the extrinsic curvature as 
%============< EQUATION >==============%
%
\begin{eqnarray}
 \gamma_{ij} & = & a^2\psi^4\tilde{\gamma}_{ij}\,, 
  \nonumber\\
 K_{ij} & = & 
  a^2\psi^4\left(\frac{1}{3}K\tilde{\gamma}_{ij}
	    +\tilde{A}_{ij}\right)\,,
	     \label{eqn:decompose-metric}
\end{eqnarray}
%======================================%
and deal with the first-order differential equations 
for $(\psi, \tilde{\gamma}_{ij})$ and $(K, \til{A}_{ij})$,
together with the Hamiltonian and momentum constraint equations. 

In addition to the Einstein equations, we also have the 
field equations for the matter sector. We note that 
because we only have a single scalar field, the field equation
is equivalent to the energy momentum conservation equation 
$\nabla_\mu T^\mu{}_\nu=0$ in the present case.

In order to solve the Einstein equations, one has to
fix the gauge condition. As for the choice of temporal gauge,
we adopt the {\it uniform Hubble slicing},
%============< EQUATION >==============%
%
\begin{equation}
 K = -3H(t), \quad H(t)\equiv \frac{\partial_t a}{a}\,. 
 \label{eqn:uniform-Hubble}
\end{equation}
%======================================%
The spatial gauge condition will be specified later,
by (\ref{eqn:additional-gauge-condition}). 

Now we employ the spatial gradient expansion. In this approach 
we assume that the characteristic length scale over which
the metric varies is much larger than the characteristic time
scale over which the metric varies. To apply it to the cosmological
situation, we introduce a flat FLRW universe ($a(t)$, $\phi_0(t)$) as a
background,\footnote{In this section, the subscript $0$ indicates
the background quantities.}
and suppose that the characteristic length scale $L$ of perturbations is
longer the Hubble length scale $1/H$ of the background, i.e., $HL\gg 1$.
Then we attach a small parameter $\epsilon$ to each spatial derivative
in the field equations and expand them in $\epsilon$.
Physically, the parameter is equivalent to the ratio of 
the Hubble radius to the length scale of perturbations, $\epsilon=1/(HL)$.

The background flat FLRW universe ($a(t)$, $\rho_0(t)$) satisfies the
Friedmann equation and the equation of motion 
%============< EQUATION >==============%
%
\begin{equation}
 H^2 = \frac{\kappa^2}{3}\rho_0, \quad 
 \frac{2}{a^3}\partial_t\left(a^3P_{0X}\partial_t\phi_0\right) -
 P_{0\phi} = 0,
 \label{eqn:background-eq}
\end{equation}
%======================================%
where $\kappa^2=8\pi G_N$, $\rho_0\equiv \rho(X_0,\phi_0)$, 
$P_{0X}\equiv P_X(X_0,\phi_0)$, 
$P_{0\phi}\equiv P_{\phi}(X_0,\phi_0)$,  and
$X_0\equiv(\partial_t\phi_0)^2$. 
Since the FLRW background must be recovered in the limit $\epsilon\to 0$,
natural assumptions on the metric are
%============< EQUATION >==============%
%
\begin{equation}
 v^i=O(\epsilon), \quad \beta^i=O(\epsilon), 
  \label{eqn:assumption-vbeta}
\end{equation}
%======================================%
and $\partial_t\tilde{\gamma}_{ij}=O(\epsilon)$. 
Actually, for this last assumption, following the
arguments in ~\cite{Sasaki:1995aw,Tanaka:2006zp,Tanaka:2007gh},
we assume a stronger condition,
%============< EQUATION >==============%
%
\begin{equation}
 \partial_t\tilde{\gamma}_{ij} = O(\epsilon^2). 
 \label{eqn:assumption-gamma}
\end{equation}
%======================================%
This corresponds to assuming the absence of any decaying modes
at leading order in the gradient expansion, namely, the absence
of spatially homogeneous anisotropy.
This is justified in most of the inflationary models in which
the number of $e$-folds of inflation $N$ is much larger than
the number required to solve the horizon and flatness problem,
$N\gg 60$.%%%%%%%%%%%%%%%%%%%%%%
%
% foot note =================================================
\footnote{Hamazaki \cite{Hamazaki:2008mh} 
solved the nonlinear equation on the leading order 
in the gradient expansion, under a more general condition, 
$\partial_t\tilde{\gamma}_{ij}=O(\delta_c)=O(\epsilon^0)$, where 
$\delta_c$ is a small parameter characterizing the amplitude
of decaying modes.}
%
%%%%%%%%%%%%%%%%%%%%%%

Applying the above assumptions to the field equations,
we can estimate the orders of magnitude of various quantities
in the gradient expansion. In summary, including the assumptions, 
we obtain the following estimates:
%============< EQUATION >==============%
%
\begin{eqnarray}
 & & 
  \psi = O(1), \quad \tilde{\gamma}_{ij} = O(1), \quad v^i=O(\epsilon),
  \quad \beta^i=O(\epsilon), 
\nonumber\\
 & & 
  \chi= O(\epsilon^2), 
\quad \tilde{A}_{ij} = O(\epsilon^2), 
  \quad \delta =O(\epsilon^2), \quad \varphi=O(\epsilon^2),  
  \quad p = O(\epsilon^2), 
\nonumber\\
 & & 
  \partial_t\tilde{\gamma}_{ij}=O(\epsilon^2), \quad
  \partial_t\psi=O(\epsilon^2), \quad
  v^i+\beta^i = O(\epsilon^3),    
\label{eqn:order-of-magnitude}
\end{eqnarray}
%======================================%
where $\chi$, $\delta$, $p$ and $\varphi$
represent the fluctuations 
in $\alpha$, $\rho$, $P$ and $\phi$, respectively, defined as 
\begin{eqnarray}
\displaystyle \chi\equiv \alpha-1\,,
\quad
\delta\equiv {\rho-\rho_0\over \rho_0}\,,
\quad
p\equiv P-P_0\,, 
\quad
\varphi\equiv\phi-\phi_0\,.
\end{eqnarray}
Note that, these fluctuations may be non-vanishing at 
leading order in the gradient expansion in general.
The advantage of the uniform Hubble slicing is that
these fluctuations all become of $O(\epsilon^2)$~\cite{Tanaka:2007gh}.
We also note that 
the form of $p$ for the scalar field system is specified by the relation
(\ref{eqn:deltaP}) as 
%============< EQUATION >==============%
%
\begin{equation}
 p = \rho_0(c_{s0}^2\delta + \Gamma_0\varphi) + O(\epsilon^4),
  \label{eqn:p-delta-pi}
\end{equation}
%======================================%
where $c_{s0}^2=P_{0X}/(2P_{0XX}X_0+P_{0X})$ and  
$\Gamma_0=(P_{0\phi}-c_{s0}^2\rho_{0\phi})/\rho_0$. 
 
We now spell out the general solution valid up through
$O(\epsilon^2)$ in the gradient expansion~\cite{Takamizu:2008ra}.
We attach the superscript $(m)$ to a quantity of $O(\epsilon^m)$.
At leading order, the only non-trivial quantities are
$\psi$ and $\tilde{\gamma}_{ij}$, which are given by
%============< EQUATION >==============%
%
\begin{equation}
 \psi = L^{(0)}(x^k) + O(\epsilon^2),
  \label{eqn:psi-leading}
\end{equation}
and 
%======================================%
\begin{equation}
 \tilde{\gamma}_{ij} = f^{(0)}_{ij}(x^k) + O(\epsilon^2),
  \label{eqn:gammatilde-leading}
\end{equation}
%======================================%
where $L^{(0)}(x^k)$ is an arbitrary function of the spatial coordinates
$\{x^k\}$ ($k=1,2,3$) and 
$f^{(0)}_{ij}(x^k)$ is a ($3\times 3$)-matrix function of the 
spatial coordinates with unit determinant, respectively.

To obtain the solution at second-order, it is convenient to
constrain the shift vector more strongly than indicated by
(\ref{eqn:order-of-magnitude}): We set
%============< EQUATION >==============%
%
\begin{equation}
 \beta^i = O(\epsilon^3). 
  \label{eqn:additional-gauge-condition}
\end{equation}
%======================================%
The choice of $\beta^i=0$ is called the {\it time-orthogonal} gauge.
We note that setting $\beta^i=O(\epsilon^3)$ (or even $\beta^i=0$ exactly)
does not fix the spatial coordinates completely~\cite{Tanaka:2006zp}.
There remains 3 gauge degrees of freedom corresponding
to the coordinate transformation,
\begin{eqnarray}
x^i\to \bar{x}^i=f^i(x^i)=O(1)\,.
\end{eqnarray}
If we fix this $O(1)$ part of the gauge, 
then the gauge is completely fixed at $O(\epsilon^2)$ accuracy.
Note also that the spatial gauge condition (\ref{eqn:additional-gauge-condition})
with $v^i+\beta^i=O(\epsilon^3)$ in (\ref{eqn:order-of-magnitude}) leads to 
the {\it comoving threading} condition: $v^i=0$ at $O(\epsilon^2)$ accuracy.

With the above choice of gauge, the general solution valid up 
to $O(\epsilon^2)$ was obtained in~\cite{Takamizu:2008ra}. It is
given by
%============< EQUATION >==============%
%
\begin{eqnarray}
 \delta & = & \frac{R^{(0)}}{2\kappa^2\rho_0a^2}
   + O(\epsilon^4), \nonumber\\
 u_i & = & \frac{1}{6\kappa^2(\rho_0+P_0)a^3}
  \partial_i 
  \left(R^{(0)} \int^t_{t_*}a(t')dt' +\tilde{C}^{(2)}\right) + O(\epsilon^5),
\nonumber\\
 \varphi & = & -\frac{\dot\phi_0}{6\kappa^2(\rho_0+P_0)a^3}
  \left(R^{(0)} \int^t_{t_*}a(t')dt' +
  \tilde{C}^{(2)}\right)  + O(\epsilon^4),  
\nonumber\\
 \chi & = & - \frac{1}{6\kappa^2(\rho_0+P_0)a^2}
  \left[
   \left(1+3c_{s0}^2 - \frac{\rho_0\Gamma_0\partial_t\phi_0}
    {(\rho_0+P_0)a} \int^t_{t_*}a(t')dt'\right) 
   R^{(0)}
   -\frac{\rho_0\Gamma_0\partial_t\phi_0}
   {(\rho_0+P_0)a}\tilde{C}^{(2)}\right] + O(\epsilon^4), 
\nonumber\\
 \psi & = & \left(L^{(0)}+L^{(2)}\right)
  \left(1+
  \frac{1}{2}\int^t_{t_*}H(t')\chi(t')dt'
\right) + O(\epsilon^4),  
\nonumber\\
 \tilde{\gamma}_{ij} & = & f^{(0)}_{ij}+f^{(2)}_{ij}
  - 2\left(
      F^{(2)}_{ij}\int^t_{t_*}\frac{dt'}{a^3(t')}\int^{t'}_{t_*}a(t'')dt''
      + C^{(2)}_{ij}\int^t_{t_*}\frac{dt'}{a^3(t')}\right)
  + O(\epsilon^4), 
\nonumber\\
 \tilde{A}_{ij} & = & \frac{1}{a^3}
  \left(F^{(2)}_{ij}\int_{t_*}^t a(t')dt' + C^{(2)}_{ij}\right) 
  + O(\epsilon^4),
  \label{eqn:general solution}
\end{eqnarray}
where the dot ($\dot{~}$) denotes $d/dt$ and
\begin{eqnarray}
 F^{(2)}_{ij}(x^k) & \equiv &\frac{1}{(L^{(0)})^4}
R_{ij}\left[(L^{(0)})^4f^{(0)}\right]
-\frac{1}{3}f^{(0)}_{ij}R\left[(L^{(0)})^4f^{(0)}\right]
\nonumber\\
&=&\frac{1}{\left(L^{(0)}\right)^4}
  \left[ \left(\tilde{R}^{(0)}_{ij}-\frac{1}{3}\tilde{R}^{(0)}f^{(0)}_{ij}\right)
   + 2\left(2\partial_i\ln L^{(0)}\partial_j\ln L^{(0)}
   - \tilde{D}^{(0)}_i\tilde{D}^{(0)}_j\ln L^{(0)}\right)\right.
\nonumber\\
 & & \left.
   - \frac{2}{3}f_{(0)}^{kl}
   \left(2\partial_k\ln L^{(0)}\partial_l\ln L^{(0)}
   - \tilde{D}^{(0)}_k\tilde{D}^{(0)}_l\ln L^{(0)} \right) f^{(0)}_{ij}
  \right],
 \label{eqn:def-F2ij}
\end{eqnarray}
%======================================%
and $\tilde{R}^{(0)}_{ij}=R_{ij}[f^{(0)}]$ and $\tilde{R}^{(0)}=R[f^{(0)}]$
are the Ricci tensor and the Ricci scalar of the $0$th-order spatial metric
$f^{(0)}_{ij}$, and $R^{(0)}=R\left[(L^{(0)})^4f^{(0)}\right]$ is 
the Ricci scalar of the $0$th-order spatial metric $(L^{(0)})^4 f^{(0)}_{ij}$.
Here note that the `constants' of integration, $L^{(2)}$ and $f^{(2)}_{ij}$,
were absorbed into $L^{(0)}$ and $f^{(0)}_{ij}$, respectively, 
in \cite{Takamizu:2008ra}, while in this paper we write them explicitly 
for later convenience. 
The choice of the initial time of integration, $t_*$,
will be discussed in the next section. 
The `constants` of integration 
$L^{(0)}$, $f^{(0)}_{ij}$, $\tilde{C}^{(2)}$ and $C^{(2)}_{ij}$ 
are not mutually independent due to the Hamiltonian and 
momentum constraints. They must satisfy 
%============< EQUATION >==============%
%
\begin{eqnarray}
 f^{(0)}_{ij} & = & f^{(0)}_{ji}, \quad \det(f^{(0)}_{ij}) = 1,
  \nonumber\\
 {C}^{(2)}_{ij} & = & C^{(2)}_{ji}, \quad f_{(0)}^{ij}C^{(2)}_{ij} = 0,
  \nonumber\\
 \left(L^{(0)}\right)^6 \partial_i\tilde{C}^{(2)} & = & 
  6 f_{(0)}^{jk}\tilde{D}^{(0)}_j
  \left[\left(L^{(0)}\right)^6{C}^{(2)}_{ki}\right], 
 \label{eqn:constraint-C2}
\end{eqnarray}
%======================================%
where $f_{(0)}^{ij}$ is the inverse matrix of $f^{(0)}_{ij}$ and
$\tilde{D}^{(0)}$ is the covariant derivative with respect
to $f^{(0)}_{ij}$. 

%======================================%

%%%%%%%%%%%%%%%%%%%%%%%%%%%%%%%%%%%%%%%%%%%%%%%%%%%%%%%
\section{Nonlinear curvature perturbation}
\label{sec:nlcurvpert}

In this section, we define a new variable that is a nonlinear
generalization of the comoving curvature perturbation up through
$O(\epsilon^2)$ in the gradient expansion. We then construct it explicitly.
We find that this newly defined variable satisfies a nonlinear 
second-order differential equation which is a generalization of
the equation for the linear comoving curvature perturbation.
In this and the following sections, we omit the subscript $0$ 
from the background quantities since there will be no danger of confusion.

\subsection{Assumptions and definitions}
\label{subsec:assumption}
As mentioned in the previous section, it is necessary to
fix the spatial gauge to fix the metric completely. 
To do so, we assume that the contribution of gravitational waves
to $\tilde\gamma_{ij}$ is negligible. In other words, we focus
on the contribution arising from the scalar-type perturbations.
Then at sufficiently late times of inflation, we may
choose the gauge in which $\tilde\gamma_{ij}$ approaches
the flat metric,
\begin{eqnarray}
\tilde{\gamma}_{ij}\to\delta_{ij}\quad (t\to\infty)\,,
\label{eq: gamma-ij t-infty}
\end{eqnarray}
where in reality the limit $t\to\infty$ may be reasonably interpreted
as an epoch close to the end of inflation.
This condition completely kills the remain 3 gauge degrees of freedom
up through $O(\epsilon^2)$ in the gradient expansion.
It may be worth noting that one may include gravitational
waves by relaxing the above condition to
\begin{eqnarray}
\partial^i(\ln \tilde{\gamma})_{ij}\to0\quad (t\to\infty)\,.
\end{eqnarray}
Imposing this condition at all times is equivalent to 
the gauge chosen in~\cite{Maldacena:2002vr}.

At leading order in the gradient expansion,
one may define the nonlinear curvature perturbation $\zeta$ by
\begin{eqnarray}
\psi^4=e^{2\zeta}\,.
\label{rel: psi vs R}
\end{eqnarray}
It is known that $\zeta$ on the uniform Hubble slices
is equal to that on the comoving ($=$ uniform scalar field) slices
at leading order in the gradient expansion~\cite{Lyth:2005fi}.
However, at second-order in the gradient expansion, 
This equivalence between the uniform Hubble slicing and
the comoving slicing breaks down. Furthermore, the very
definition of the nonlinear curvature perturbation~(\ref{rel: psi vs R})
becomes inadequate as seen below.

Let us derive the relation between $\zeta_H$ and $\zeta_c$ to 
$O(\epsilon^2)$, where and in what follows we use the subscripts
$H$ and $c$ to denote the quantities evaluated on the uniform Hubble 
and comoving slices, respectively. 
We have obtained the solution~(\ref{eqn:general solution})
in the {\it uniform Hubble slicing} (temporal), {\it time-orthogonal} 
(spatial) gauge, 
\begin{eqnarray}
K=-3H(t)\,,\quad \beta^i(t,x^i)=0\,.
\end{eqnarray}
In general, we have to consider a nonlinear transformation
between different time slices.
However, since $\varphi_H=O(\epsilon^2)$, the transformation to 
the comoving slicing, $\varphi_c=0$,
happens to be just like a linear gauge transformation.
The comoving curvature perturbation is obtained as
\begin{eqnarray}
{\zeta}_c={\zeta}_H-\frac{H}{\dot{\phi}}\varphi_H+O(\epsilon^3). 
\label{def: comoving nonlinear curvature perturb}
\end{eqnarray}
Then the general solution for ${\zeta}_c$ valid up to 
the $O(\epsilon^2)$ in the gradient expansion 
in the {\it comoving slicing, time-orthogonal} gauge,
\begin{eqnarray}
\varphi_c(t,x^i)=\beta_c^i(t,x^i)=0, 
\label{def: comoving-time-orth}
\end{eqnarray}
is written by the solutions of $\psi$ and $\varphi$ in 
(\ref{eqn:general solution}), whose explicit form will 
be given in the next subsection. Here note that $\tilde\gamma_{ij}$
remains the same at $O(\epsilon^2)$ accuracy under the change of 
time-slicing from the uniform Hubble slicing to the comoving slicing,
\begin{eqnarray}
\tilde\gamma_{H,ij}=\tilde\gamma_{c,ij}+O(\epsilon^4)\,.
\end{eqnarray}

Now we turn to the problem of properly defining a nonlinear
curvature perturbation to $O(\epsilon^2)$ accuracy.
Let us denote the linear curvature perturbation on 
comoving slices by ${\cal R}^{\rm Lin}_c$. 
%%%%%%%%%%%%%%%%%%%%%%%%%%%%%%%%%%%%%%%%%%%%%%
%
%nonlinear H_T
%
%%%%%%%%%%%%%%%%%%%%%%%%%%%%%%%%%%%%%%%%%%%%%%
In the linear limit, the variable ${\zeta}_c$ reduces to 
${\cal R}^{\rm Lin}_c$ at leading order in the gradient expansion, 
but not at second-order.
The comoving curvature perturbation in the linear limit
is given by
\begin{eqnarray}
{\cal R}^{\rm Lin}=\left(H^{\rm Lin}_L+{H^{\rm Lin}_T\over 3}\right)Y,
\label{def: linear curvature}
\end{eqnarray}
where, following the notation in \cite{Kodama:1985bj},
the spatial metric in the linear limit is expressed as 
\begin{eqnarray}
\gamma_{ij}=a^2(\eta)(\delta_{ij}
+2 H^{\rm Lin}_LY \delta_{ij}+2 H^{\rm Lin}_T Y_{ij})\,,
\end{eqnarray}
with $Y$ being scalar harmonics with eigenvalue $k^2$ satisfying
\begin{eqnarray}
(\Delta + k^2)Y_{\bm{k}}=0\,,
\end{eqnarray}
and 
\begin{eqnarray}
Y_{ij}=k^{-2} \left[
\partial_i \partial_j -{1\over 3} \delta_{ij} \Delta \right] Y\,.
\label{eq: scalar harmonics}
\end{eqnarray}
These expressions in linear theory correspond to the metric components in our
notation as
\begin{eqnarray}
{\zeta}=
 H^{\rm Lin}_L Y,~~\til{\gamma}_{ij} = \delta_{ij} + 2H^{\rm Lin}_T Y_{ij},
\label{def: notition linear quantity}
\end{eqnarray}
where we have used the definition of $\zeta$ given by (\ref{rel: psi vs R}). 
In the above we have omitted the eigenvalue indices and the sum over 
the eigen modes for notational simplicity, but the existence of
it is implicitly assumed. For example,
$Q=qY$ means
\begin{eqnarray}
Q(t,x^i)=\sum_{\bm k}q_{\bm k}(t)Y_{\bm k}(x^i)\,.
\end{eqnarray}

Thus to define a nonlinear generalization of the linear curvature
perturbation~(\ref{def: linear curvature}), we need nonlinear generalizations
of $H_LY$ and $H_TY$. Our nonlinear ${\zeta}$ is an
apparent natural generalization of $H^{\rm Lin}_LY$,
\begin{eqnarray}
H_LY=\zeta\,.
\end{eqnarray}
As for $H_TY$, however, the generalization is non-trivial.
Nevertheless, because it is the $O(\epsilon^2)$ part of $\tilde\gamma_{ij}$,
the correspondence may be assumed to be linear. Hence, by introducing
the inverse Laplacian operator $\Delta^{-1}$ on the flat background,
we define the nonlinear generalization of $H_TY$ as
\begin{eqnarray}
H_TY=E\equiv
-\frac{3}{4}\Delta^{-1}\left[\partial^i \psi^{-6}
\partial^j \psi^{6}(\ln \tilde{\gamma})_{ij} \right].
\label{def: nonlinear HT0}
\end{eqnarray}
With these definitions of $H_LY$ and $H_TY$, we can define the nonlinear 
curvature perturbation valid up through $O(\epsilon^2)$ as
\begin{eqnarray}
{\cal R}^{\rm NL}\,\equiv\, {\zeta}\,+\,{E\over 3}\,.
\label{def0: nonlinear variable zeta}
\end{eqnarray}

As clear from (\ref{def: nonlinear HT0}), finding $H_TY$ generally
requires a spatially non-local operation. However, as we shall see 
in the next subsection, in the comoving slicing, time-orthogonal gauge
with the asymptotic condition on the spatial 
coordinates~(\ref{eq: gamma-ij t-infty}), we find it
is possible to obtain the explicit expression for the nonlinear
version of $H_TY$ from our solution (\ref{eqn:general solution}) 
without any non-local operation.

%%%%%%%%%%%%%%%%%%%%%%
%
%
%---------------IIB---------------
%
%
%%%%%%%%%%%%%%%%%%%%%%%%%%%%
\subsection{Explicit expression}
\label{subsec:explicit}

First we derive the explicit expression of $\zeta_c$.
 Using (\ref{def: comoving nonlinear curvature perturb}) with
 (\ref{rel: psi vs R}), it is obtained from the general 
solution (\ref{eqn:general solution}) as 
\begin{eqnarray}
{\zeta}_c=\ell^{(0)}+\tilde{\ell}^{(2)}+{f}_K(t)\,K^{(2)}+
{f}_C (t)\, \tilde{C}^{(2)}+O(\epsilon^4),
\label{sol: general sol R}
\end{eqnarray}
where $\ell^{(0)}=2\ln L^{(0)}$ and $\ell^{(2)}=2L^{(2)}/L^{(0)}$,
The functions $f_K(t)$ and $f_C(t)$ are defined by
\begin{eqnarray}
&&{f}_K(t)=
-{1\over 3\kappa^2}\left[\int_{t_*}^{t} 
{H\over 2(\rho+P)a^2} \left(1+3 c_s^2-{\rho 
\Gamma \dot{\phi}\over (\rho+P) a}
\int_{t_*}^{t'} a(t'') dt''\right) dt'
-{H\over 2(\rho+P)a^3}\int_{t_*}^{t} a(t') dt'\right],
\label{eq: f-K t} \\
&&{f}_C(t)
={1\over 3\kappa^2}
\left[\int_{t_*}^{t} {H\rho \Gamma\dot{\phi}\over 2(\rho+P)^2a^3}dt'
 + {H\over 2(\rho+P)a^3}\right].
\label{eq: f-C t}
\end{eqnarray}
The terms $K^{(2)}$ and $\tilde{C}^{(2)}$ are 
spatial functions, and the former is the Ricci scalar of the 
$0$th-order spatial metric,
\begin{eqnarray}
K^{(2)}[\ell^{(0)}]=&&R\left[(L^{(0)})^4f^{(0)}\right]=
-{8\Delta L^{(0)}\over (L^{(0)})^5}
\nonumber\\
=&&
{-2 (2 \Delta \ell^{(0)}+\delta^{ij}\partial_i \ell^{(0)}
\partial_j \ell^{(0)})e^{-2 \ell^{(0)}}}\,,
\label{def: K2}
\end{eqnarray}
while the latter $\tilde{C}^{(2)}$ is arbitrary for the moment.

At leading order in the gradient expansion, $\ell^{(0)}$
is the conserved comoving curvature perturbation, equivalent to 
the fluctuation in the number of $e$-folds $\delta N$ from some
final uniform density (or comoving) hypersurface to the
initial flat hypersurface (on which $\zeta=0$) at $t=t_*$,
\begin{eqnarray}
\ell^{(0)}=\delta N(t_*,x^i)\,.
\label{eq: delta-N-u0}
\end{eqnarray} 
However, at second-order this is no longer the case, since
$\zeta_c(t_*)$ is not equal to $\zeta_c(\infty)$ in general as
clear from (\ref{sol: general sol R}). The `constants'
of integration $\ell^{(2)}=2L^{(2)}/L^{(0)}$ and $\tilde{C}^{(2)}$ 
at $O(\epsilon^2)$ characterize the total time variation of $\zeta_c$
 from $t=t_*$ to $t=\infty$.

%%%%%%%%%%%%%%%%%%%%%%%%%%%%%%%%%%%%%%%%%%%%%%
%
%nonlinear H_T in explicit form
%
%%%%%%%%%%%%%%%%%%%%%%%%%%%%%%%%%%%%%%%%%%%%%%

Next we consider the explicit expression of $\tilde\gamma_{ij}$
in the comoving slicing, time-orthogonal gauge.
As mentioned in the previous subsection, $\tilde\gamma_{ij}$ is
the same for both this gauge and the original uniform Hubble, time-orthogonal
gauge. Hence it is the one given in (\ref{eqn:general solution}),
\begin{eqnarray}
\til{\gamma}_{ij}=f^{(0)}_{ij}+f^{(2)}_{ij}-2F_{ij}^{(2)}A(t)
-2C_{ij}^{(2)}B(t)+O(\epsilon^4),
\label{eq: til-gamma-ij}
\end{eqnarray}
where we have introduced the integrals,
\begin{eqnarray}
A(t)=\int_{t_*}^{t}
{dt'\over a^3(t')}\int_{t_*}^{t'} a(t'') dt''\,,
\quad
B(t)=\int_{t_*}^t {dt'\over a^3(t')}\,. 
\label{def: integral-A-B}
\end{eqnarray}
The time-independent terms $f^{(0)}_{ij}$ and $f^{(2)}_{ij}$
are determined from the condition (\ref{eq: gamma-ij t-infty}) as
\begin{eqnarray}
&&f_{ij}^{(0)} = \delta_{ij}\,,\nonumber\\
&&f_{ij}^{(2)} =2 F^{(2)}_{ij}A(\infty)+ 2 C^{(2)}_{ij}B(\infty)\,.
\label{def: f2-ij}
\end{eqnarray}
$F_{ij}^{(2)}$ is given by (\ref{eqn:def-F2ij}) with
 $\ell^{(0)}=2 \ln L^{(0)}$,
\begin{eqnarray}
F_{ij}=e^{-2\ell^{(0)}} \Bigl[\partial_i \ell^{(0)}\partial_j \ell^{(0)}
-\partial_i \partial_j
\ell^{(0)} -{1\over 3} \left(\partial_k \ell^{(0)} \partial^k \ell^{(0)}
 -\Delta \ell^{(0)}\right)
\delta_{ij}\Bigr],
\label{def: F-ij}
\end{eqnarray}
and $C^{(2)}_{ij}$ satisfies the constraint equation, 
\begin{eqnarray}
e^{3\ell^{(0)}} \partial_i\tilde{C}^{(2)} =  
  6 \partial^j\left(e^{3\ell^{(0)}}C^{(2)}_{ji}\right).
\label{eq: const-C^2-ij}
\end{eqnarray}

Now, from the above expression of $\tilde\gamma_{ij}$, we derive 
the explicit expression of $E=H_TY$. The definition
of $E$ in (\ref{def: nonlinear HT0}) with $\ell^{(0)}=2 \ln L^{(0)}$ gives 
\begin{eqnarray}
E_c
=-{3\over 4}
\Delta^{-1}\left[\partial^i e^{-3 \ell^{(0)}}\partial^j e^{3\ell^{(0)}}
 (\tilde{\gamma}_{ij}-\delta_{ij})\right]\,.
\label{def: nonlinear HT}
\end{eqnarray}
Hence we first need to evaluate the following expressions:
\begin{eqnarray}
\partial^i\left[e^{-3 \ell^{(0)}}\partial^j\left(e^{3\ell^{(0)}}
 F^{(2)}_{ij}\right)\right]\,,
\quad
\partial^i\left[e^{-3 \ell^{(0)}}\partial^j\left( e^{3\ell^{(0)}}
 C^{(2)}_{ij}\right)\right]\,.
\end{eqnarray}
The latter can be immediately evaluated from the
constraint (\ref{eq: const-C^2-ij}) to be
\begin{eqnarray}
\Delta\,\tilde{C}^{(2)}
 ={6} \partial^i\left[e^{-3 \ell^{(0)}}\partial^j\left( e^{3\ell^{(0)}}
 C^{(2)}_{ij}\right)\right]\,.
\end{eqnarray}
As for the former, recalling that $F^{(2)}_{ij}$ is
the traceless part of the Ricci tensor of the metric
$e^{2\ell^{(0)}}f^{(0)}_{ij}$, we can verify
\begin{eqnarray}
\Delta\,K^{(2)}
={6}\partial^i\left[e^{-3 \ell^{(0)}}\partial^j\left(e^{3\ell^{(0)}}
 F^{(2)}_{ij}\right)\right]\,.
\end{eqnarray}
Thus by substituting (\ref{eq: til-gamma-ij}) into 
(\ref{def: nonlinear HT}) with (\ref{def: f2-ij}), we obtain
\begin{eqnarray}
E_c=3H^{(2)} +{K^{(2)}\over 4}A(t)+{\tilde{C}^{(2)}\over 4}B(t)
+O(\epsilon^4),
\label{eq: HT}
\end{eqnarray}
where we have defined 
\begin{eqnarray}
H^{(2)}=-{1\over 12}\left(K^{(2)} A(\infty)+\tilde{C}^{(2)}B(\infty)\right).
\label{def: H2-infty}
\end{eqnarray}

Of course, the linear limit of $E_c=H_TY$ found above
reduces consistently to $H_T^{\rm Lin}Y$.
Setting
\begin{eqnarray}
C_{ij}^{(2)}\approx
{1\over 4}\Delta^{-1}(\partial_i \partial_j -{1\over 3} \delta_{ij} 
\Delta)\tilde{C}^{(2)}\equiv-\frac{1}{4}\tilde{c}^{(2)}Y_{ij}\,,
\end{eqnarray} 
which is consistent with the constraint (\ref{eq: const-C^2-ij})
in the linear limit, and
\begin{eqnarray}
F_{ij}^{(2)}\approx -\left(\partial_i \partial_j-{1\over 3}
 \delta_{ij} \Delta\right) \ell^{(0)}
\equiv k^2 u^{(0)}Y_{ij}\,,
\end{eqnarray}
we find 
\begin{eqnarray}
{H^{\rm Lin}_T}Y=\left(3H^{(2)}_{\rm Lin}+
{k^2 u^{(0)}} A(t)+\frac{\tilde{c}^{(2)}}{4}B(t)\right)Y
+O(\epsilon^4),
\label{eq: linear-HT}
\end{eqnarray}
with 
\begin{eqnarray}
H^{(2)}_{\rm Lin}=-{1\over 3}\left(k^2  u^{(0)}
A(\infty)+\frac{\tilde{c}^{(2)}}{4}B(\infty)\right).
%\label{def: H2-infty}
\end{eqnarray}

Finally, we obtain the nonlinear curvature perturbation ${\cal R}_c^{\rm NL}$ 
defined by (\ref{def0: nonlinear variable zeta})
in the comoving slicing, time-orthogonal gauge as
\begin{eqnarray}
{\cal R}_c^{\rm NL}(t)= &&{\zeta}_c+{E_c\over 3}
\nonumber\\
=&& \ell^{(0)}+\tilde{\ell}^{(2)}+K^{(2)}{f}_K(t)
+\tilde{C}^{(2)}f_C(t)
+H^{(2)} +{K^{(2)}\over 12}A(t)+{\tilde{C}^{(2)}\over 12}B(t)\,,
\label{def: nonlinear variable zeta}
\end{eqnarray}
where the functions ${f}_K(t)$, ${f}_C(t)$, $A(t)$ and $B(t)$ 
have been defined in (\ref{eq: f-K t}), (\ref{eq: f-C t}) 
and (\ref{def: integral-A-B}).
This will be the basic variable to be matched to
the solution in $n$-th-order perturbation theory ($n\geq 1$).
The determination of 
$\ell^{(0)}$, $\tilde{\ell}^{(2)}$ and $\tilde{C}^{(2)}$ 
by this matching will be discussed in the next section. 
%%%%%%%%%%%%%%%%%%%%%%%%%%%%%%%%%%%%%%
%
%
\subsection{$t_*$-shift}
\label{subsec:tkshift}
%
%
%%%%%%%%%%%%%%%%%%%%%%%%%%%%%%%%%%%%%%%%%%
Here, we investigate the dependence of our general nonlinear solution
${\cal R}_c^{\rm NL}(t)$ on the initial time $t_*$ which 
appears in the integrals in the functions
 ${f}_K(t)$, ${f}_C(t)$, $A(t)$ and $B(t)$.
Apparently the solution (\ref{def: nonlinear variable zeta})
depends on $t_*$ if $\tilde{\ell}^{(2)}$ and $\tilde{C}^{(2)}$ are
$t_*$-independent. However, if we match the perturbative
solution whose initial condition has been fixed 
deep inside the horizon to our nonlinear solution on superhorizon
scales ${\cal R}_c^{\rm NL}(t)$, it should not depend on the choice
of $t_*$. This implies that the `constants' of integration (purely
spatial functions) $\tilde{\ell}^{(2)}$ and $\tilde{C}^{(2)}$ must
depend on $t_*$ in such a way to cancel the $t_*$-dependence of
the temporal functions ${f}_K(t)$, ${f}_C(t)$, $A(t)$ and $B(t)$.
In particular, this implies that ${\cal R}_c^{\rm NL}$
must be invariant under an infinitesimal shift of $t_*$,
\begin{eqnarray}
t_*\to t_*+\delta t\,.
\label{def: trans-tk}
\end{eqnarray}
Thus the invariance of ${\cal R}_c^{\rm NL}$ on the variation of $t_*$
gives consistency conditions that must be satisfied in the matching
formulas derived in the next section.

Let us consider the variation of ${\cal R}_c^{\rm NL}$ with 
respect to this boundary time $t_*$ in the integrals,
\begin{eqnarray}
\delta_{t_*} {\cal R}_c^{\rm NL}={\cal R}_c^{\rm NL}(t_*+\delta t)-
{\cal R}_c^{\rm NL}(t_*), 
\end{eqnarray}
where $\delta_{t_*}$ is the variation with respect to $t_*$. 
From (\ref{eq: f-K t}) and (\ref{eq: f-C t}), we find 
\begin{eqnarray}
&&\delta_{t_*} {f}_K(t)=
-\left({f}_C(t) a(t_*)-{g_1(t_*)\over 3\kappa^2}\right)\delta t
+O(\delta t^2),\nonumber\\
&&\delta_{t_*} {f}_C(t)=
-{g_2(t_*)\over 3 \kappa^2} \delta t+O(\delta t^2), 
\end{eqnarray}
where we have defined 
\begin{eqnarray}
g_1(t)\equiv {H(1+3c_s^2)\over 2(\rho+P)a^2}\bigg|(t)\,,
\quad
g_2(t)\equiv {H\rho \Gamma \dot{\phi}\over 2(\rho+P)^2 a^3}\bigg|(t)\,.
\label{def: fun-g1-g2}
\end{eqnarray}
Similarly, we obtain from (\ref{def: integral-A-B}),
\begin{eqnarray}
&&\delta_{t_*} A(t)= -a(t_*) B(t) \delta t+O(\delta t^2),
\nonumber\\
&&\delta_{t_*} B(t)= -{\delta t\over a^3(t_*)}+O(\delta t^2).
\end{eqnarray}
Hence the variation of ${\cal R}_c^{\rm NL}$ is given by 
\begin{eqnarray}
\delta_{t_*}{\cal R}_c^{\rm NL}(t)
=&&-\left({f}_C(t) a(t_*)
-{g_1(t_*)\over 3\kappa^2}\right)K^{(2)}\, \delta t 
-{g_2(t_*)\over 3 \kappa^2} \tilde{C}^{(2)} \delta t\nonumber\\
&&-a(t_*) B(t) {K^{(2)}\over 12}\delta t -{1\over a^3(t_*)}
{\tilde{C}^{(2)}\over 12}\delta t 
+O(\delta t^2)\,.
\end{eqnarray}
Then requiring that ${\cal R}_c^{\rm NL}$ be invariant fixes
how $\tilde{\ell}^{(2)}$ and $\tilde{C}^{(2)}$
 (and hence $H^{(2)}$) should transform under this variation:
\begin{eqnarray}
\tilde{C}^{(2)}&&\to\  \tilde{C}^{(2)}+K^{(2)} a(t_*) \delta t\,,
\nonumber\\
\tilde{\ell}^{(2)}&&\to\  \tilde{\ell}^{(2)}+{1\over 3\kappa^2}
\left[-g_1(t_*)K^{(2)}+g_2(t_*)\tilde{C}^{(2)}\right]\delta t\,,
\nonumber\\
H^{(2)}&&\to\  H^{(2)}+{\tilde{C}^{(2)}{\delta t}\over 12a^3(t_*)}\,.
\label{eq: trans-C-L-tk}
\end{eqnarray}

Before closing this subsection, we mention that
not only ${\cal R}_c^{\rm NL}$ but also 
each of ${\zeta}_c$ and $E_c$ is $t_*$-independent as well.
This is a reflection of the fact that they are all gauge-invariant
because the gauge has been completely fixed.

%%%%%%%%%%%%%%%%%%%%%%%%%%%%%%%%%%%%%%%%%%%%%
% 
%
\subsection{Second-order differential equation}
\label{subsec:nleq}
%
%
%%%%%%%%%%%%%%%%%%%%%%%%%%%%%%%%%%%%%%%%%%%%
In this subsection, we derive a nonlinear second-order 
differential equation that ${\cal R}_c^{\rm NL}$ satisfies at
$O(\epsilon^2)$ accuracy.
For this purpose, we rewrite the integrals in the
functions $f_K(t)$ and $f_C(t)$ defined in
 (\ref{eq: f-K t}) and (\ref{eq: f-C t}) in terms of 
the quantity $z$ commonly used in the literature~\cite{Mukhanov:1990me},
\begin{eqnarray}
z={{a\over H}\left(\rho+P \over c_s^2\right)^{1\over 2}}\,.
\label{def: variable-z}
\end{eqnarray}
We also introduce the conformal time $\eta$,
\begin{eqnarray}
d\eta={dt\over a(t)}\,,
\label{def: conformal time}
\end{eqnarray}
and use $t$ and $\eta$ interchangeably.

In Appendix \ref{apx:formulas}, using the background equations,
we derive formulas that are used to change the forms of
the functions $f_K(t)$ and $f_C(t)$. Using these formulas, 
we obtain
\begin{eqnarray}
{f}_K(\eta)&=&-{1\over 4}\left\{
{1\over 3} \int_{\eta_*}^{\eta}
{d\eta'\over a^2(\eta')}\int_{\eta_*}^{\eta'} a^2(\eta'') d\eta'' 
+{2 a_*\over \kappa^2 H_* }\int_{\eta_*}^{\eta}
{d\eta'\over z^2(\eta')}
+\int_{\eta_*}^{\eta} 
{d\eta'\over z^2(\eta')}
\int_{\eta_*}^{\eta'} z^2 c_s^2 (\eta'') d\eta'' 
\right\},\nonumber\\
{f}_C(\eta)&=&{H_*\over 6 \kappa^2 (\rho+P)_*a_*^3}
-{1\over 12}\int_{\eta_*}^\eta {d\eta'\over a^2(\eta')} 
+\int_{\eta_*}^{\eta}{d\eta'\over 2\kappa^2 z^2(\eta')}\,,
\end{eqnarray}
where the subscript $*$ indicates the quantity evaluated
at $t=t_*$ (or $\eta=\eta_*$).
Further, in the analysis below, it is useful to adopt the 
same notation for the integrals in the above equations
as the one used in linear theory by Leach et al.~\cite{Leach:2001zf} 
and re-express them as
\begin{eqnarray}
{f}_K(\eta)&=&-{1\over 12} A(\eta) 
-{1 \over 6 \kappa^2 H_*^2 z_*^2 }\bigl[D_*-D(\eta)\bigr]-{1\over 4}
\bigl[F_*-F(\eta)\bigr]\,,
\nonumber\\
{f}_C(\eta)&=&{H_*\over 6 \kappa^2 (\rho+P)_*a_*^3}-{1\over 12}
B(\eta)+{1\over 6\kappa^2 a_* H_* z_*^2} \bigl[D_*-D(\eta)\bigr]\,,
\end{eqnarray}
where
\begin{eqnarray}
D(\eta)=3{\cal H}_* \int_{\eta}^{0}{z^2(\eta_*)\over z^2(\eta')}d\eta'\,, 
\quad
F(\eta)=\int_{\eta}^{0} 
{d\eta'\over z^2(\eta')}
\int_{\eta_*}^{\eta'} z^2 c_s^2 (\eta'') d\eta''\,,
\label{def: integral-D-F}
\end{eqnarray}
Here $D_*=D(\eta_*)$, $F_*=F(\eta_*)$
and ${\cal H}_*$ denotes the conformal Hubble parameter
 ${\cal H}=d\ln a/d\eta$ at $\eta=\eta_*$.
The functions $A(\eta)$ and $B(\eta)$ are the same as $A(t)$ and $B(t)$
defined in (\ref{def: integral-A-B}) except that they are now 
implicitly assumed to be functions of the conformal time.
Note that $t\to\infty$ corresponds to $\eta\to0$ in the conformal time.
Thus the functions $D$ and $F$ vanish asymptotically at late times,
$D(0)=F(0)=0$. Here it is important to note that the function $D(\eta)$ is
the decaying mode in the long-wavelength limit (i.e., leading order in
the gradient expansion) in linear theory,
and $F(\eta)$ is the $k^2$ correction to the growing (i.e., constant) mode,
\begin{eqnarray}
D''+2\frac{z'}{z}D'=0\,,
\quad
F''+2\frac{z'}{z}F'+c_s^2=0\,,
\label{eq:DFmeaning}
\end{eqnarray}
where the growing mode is assumed to have the form $1+k^2F(\eta)+O(k^4)$,
and the prime denotes the conformal time derivative, ${~}'=d/d\eta$.

Let us first recapitulate the solution
${\zeta}_c$ given by (\ref{sol: general sol R}) 
\begin{eqnarray}
{\zeta}_c(\eta)=\ell^{(0)}+\tilde{\ell}^{(2)}+{f}_K(\eta)\ K^{(2)}
+{f}_C (\eta)\  \tilde{C}^{(2)}+O(\epsilon^4)\,,
\label{sol: nonlinear Rc}
\end{eqnarray}
and $E_c$ given by (\ref{eq: HT}),
\begin{eqnarray}
E_c=3H^{(2)} +{K^{(2)}\over 4}A(\eta)+{\tilde{C}^{(2)}\over 4}B(\eta)
+O(\epsilon^4)\,.
\label{eq: HTeta}
\end{eqnarray}
Adding these two and using the new expressions for $f_K(\eta)$
and $f_C(\eta)$, we find that the functions $A(\eta)$ and $B(\eta)$
cancel out to yield
%
%  main zeta
%
%%%%%%%%%%%%%%%%%%%%%%%%%%%%%%%%%%%%%%%%%%%%%%%
\begin{eqnarray}
{\cal R}_c^{\rm NL}(\eta)=&&\zeta_c +\frac{E_c}{3}
\nonumber\\
=&&\ell^{(0)}+\ell^{(2)}+H^{(2)}+
{1\over 4}\bigl[F(\eta)-F_*\bigr] 
K^{(2)}+ \bigl[D(\eta)-D_*\bigr] C^{(2)}+O(\epsilon^4), 
\label{eq: sol-tild-zeta}
\end{eqnarray}
%%%%%%%%%%%%%%%%%%%%%%%%%%%%%%%%%%%%%%%%%
%
%
%
where $H^{(2)}$ is given by (\ref{def: H2-infty}),
and we have introduced the new spatial functions
$\ell^{(2)}$ and $C^{(2)}$ by 
\begin{eqnarray}
\ell^{(2)}&=&{H_*\,\tilde{C}^{(2)}\over 6 \kappa^2 (\rho+P)_*a_*^3}
 + \tilde{\ell}^{(2)}, \nonumber\\
C^{(2)}&=&-{1\over 6\kappa^2 a_* H_* z_*^2} 
\Bigl[\tilde{C}^{(2)} -{a_*\over H_*}K^{(2)} \Bigr]\,.
\end{eqnarray}
This is one of our main results. The solution turns out to have
a very simple form. In fact, as noted in the previous paragraph,
the functions $D(\eta)$ and $F(\eta)$ have special meanings
in linear theory: $D(\eta)$ is the decaying mode at leading order 
in the gradient expansion and $F(\eta)$ is the $O(\epsilon^2)$ 
correction to the growing mode, as shown in (\ref{eq:DFmeaning}).
This implies that, within the current accuracy of the gradient expansion,
our solution ${\cal R}_c^{\rm NL}$ satisfies the nonlinear
second-order differential equation,
%%%%%%%%%%%%%%%%%%%
%
%
%%%%%%%%%%%%%%%%%%%%%%%%%
\begin{eqnarray}
{{\cal R}_c^{\rm NL}}''+2 {z'\over z} 
{{\cal R}_c^{\rm NL}}' +{c_s^2\over 4} K^{(2)}[\,
{\cal R}_c^{\rm NL}\,]=O(\epsilon^4)\,,
\label{eq: basic eq for NL}
\end{eqnarray}
where $K^{(2)}[X]$ is the Ricci scalar of the metric obtained 
by replacing $\ell^{(0)}$ with $X$ in (\ref{def: K2}).
In the linear limit, it reduces to the well-known equation
for the curvature perturbation on comoving hypersurfaces,
\begin{eqnarray}
{{\cal R}^{\rm Lin}_c}''+2{z'\over z} {{\cal R}^{\rm Lin}_c}'
-c_s^2\,\Delta[{\cal R}^{\rm Lin}_c]=0\,.
\label{eq: linear eq R}
\end{eqnarray}

Equation~(\ref{eq: basic eq for NL}) may be regarded as the master
equation for nonlinear superhorizon curvature perturbations in 
second-order in the gradient expansion. It should be, however, used with caution.
For example, since it is derived under the assumption that the
decaying mode is absent at leading order in the gradient expansion,
a decaying mode solution obtained from the above equation with
$O(\epsilon^2)$ corrections cannot be justified.
Nevertheless, it would be interesting to investigate if a solution
to (\ref{eq: basic eq for NL}) with the right-hand side set to exactly
zero can actually be a useful approximation to a full nonlinear solution
on the Hubble horizon scales or even on scales somewhat smaller than
the Hubble radius.
%%%%%%%%%%%%%%%%%%%%%%%%%%%%%%%%%%%%%%%%%%%%%%%%
%
%
\section{Matching condition}
\label{sec:matching}
%
%
%%%%%%%%%%%%%%%%%%%%%%%%%%%%%%%%%%%%%%%%%%%%
The general solution~(\ref{eq: sol-tild-zeta})
for the nonlinear curvature perturbation on comoving slices 
${\cal R}_c^{\rm NL}$ has three arbitrary 
spatial functions, $\ell^{(0)}$, $\ell^{(2)}$ and $C^{(2)}$.
Note, however, that the number of physical degrees of freedom are 
two since $\ell^{(2)}$ can be absorbed into $\ell^{(0)}$.
This is consistent with the fact that ${\cal R}_c^{\rm NL}$
satisfies the second-order differential equation~(\ref{eq: basic eq for NL}),
or the fact that scalar-type perturbation has only a single 
field degree of freedom in the Lagrangian formalism.

Physically these undetermined `constants' of integration
must be determined by the initial condition at a sufficiently
early time when the scale of interest is well inside the Hubble horizon.
There since the gradient expansion is not applicable at all, 
we have to resort to the standard perturbation theory.
Let us assume that we have obtained the $n$-th-order perturbation
solution under an appropriate initial condition. Let us denote
this perturbative solution by ${\cal R}^{\rm pert}_c$.
Introducing a small expansion parameter $\delta$
(not to be confused with the density perturbation)
that characterizes the amplitude of perturbation, we may write
\begin{eqnarray}
{\cal R}^{\rm exact}_c(\eta)={\cal R}^{\rm pert}_c(\eta)
+O(\delta^{n+1})\,,
\end{eqnarray}
where ${\cal R}^{\rm exact}_c$ is the exact solution.
Our task is
to match this perturbative solution to our nonlinear
 solution on superhorizon scales where the accuracy of 
the gradient expansion to second-order is sufficient.

In this section, we perform this matching at $t=t_*$ or $\eta=\eta_*$.
We denote the $n$-th-order perturbative solution by
${\cal R}^{\rm pert}_c$. We choose the matching time
such that the characteristic comoving scale of our interest $k$
crossed the horizon about one expansion time before.
That is, we are interested in wavenumbers that satisfy
\begin{eqnarray}
\left(\frac{k}{{\cal H}_*}\right)^2=
\left(\frac{k}{a_*H_*}\right)^2=(k\eta_*)^2\ll1\,.
\label{def: horizon cross}
\end{eqnarray} 

At and around the epoch $\eta=\eta_*$, we assume that both
${\cal R}^{\rm pert}_c$ and ${\cal R}^{\rm NL}_c$ are
reasonably accurate approximations to the exact solution.
Namely, for some finite range of time interval
around $\eta=\eta_*$, we have
\begin{eqnarray}
{\cal R}^{\rm pert}_c(\eta)
={\cal R}^{\rm exact}_c(\eta)+O(\delta^{n+1})
={\cal R}^{\rm NL}_c(\eta)+O(\epsilon^4,\delta^{n+1})\,.
\label{eq:approx}
\end{eqnarray}
This implies that ${\cal R}^{\rm pert}_c(\eta)$ satisfies the same
second-order differential equation as (\ref{eq: basic eq for NL})
up to the error of $O(\epsilon^4,\delta^{n+1})$.

%%%%%%%%%%%%%%%%%%%%%%%%%%%%%%
%
%
\subsection{General formalism}
%
%
%%%%%%%%%%%%%%%%%%%%%%%%%%%%%%%%%%
Since ${\cal R}_c^{\rm NL}$ satisfies a second-order
differential equation, the solution is completely determined
once its value and the time-derivative are given at a time.
Thus, assuming we know the $n$-th-order perturbative solution, 
the matching condition at $\eta=\eta_*$ is given by 
\begin{eqnarray}
{\cal R}_c^{\rm NL}\big|_{\eta=\eta_*}&&=\,{\cal R}_c^{\rm pert}
\big|_{\eta=\eta_*}+O(\epsilon^4,\delta^{n+1}),
\nonumber\\
({{\cal R}_c^{\rm NL}})'\big|_{\eta=\eta_*}&&=\,
({\cal R}_c^{\rm pert})'\big|_{\eta=\eta_*}+O(\epsilon^4,\delta^{n+1})\,.
\label{eq: matching cond}
\end{eqnarray}
The first condition of (\ref{eq: matching cond}) leads to 
\begin{eqnarray}
\ell^{(0)}+\ell^{(2)}+H^{(2)}={\cal R}_c^{\rm pert}(\eta_*)
+O(\epsilon^4,\delta^{n+1})\,,
\label{eq: matching cond1}
\end{eqnarray}
and the second condition of (\ref{eq: matching cond}) gives 
\begin{eqnarray}
C^{(2)}=-{({{\cal R}_c^{\rm pert}})'(\eta_*)\over 3 {\cal H}_*}
+O(\epsilon^4,\delta^{n+1})\,.
\label{eq: matching cond2}
\end{eqnarray} 
Note that the above equation means 
$({{\cal R}_c^{\rm pert}})'=O(\epsilon^2)$.
This is because we have assumed that there is no decaying mode 
at leading order, $O(\epsilon^0)$, in the gradient expansion.

Using the matching conditions (\ref{eq: matching cond1}) and 
(\ref{eq: matching cond2}), we obtain
%%%%%%%%%%%%%%%%%%%%%%%%%%%%%%%%%%%%%%%%%%%%%%%
%
% matching solution
%
%
\begin{eqnarray}
{\cal R}_c^{\rm NL}(\eta)
={\cal R}_c^{\rm pert}(\eta_*)+
\Bigl[D_*-D(\eta)\Bigr]{({{\cal R}_c^{\rm pert}})'(\eta_*)\over 3{\cal H}_*} 
+{1\over 4}\Bigl[F(\eta)-F_*\Bigr]K^{(2)}[{\cal R}_c^{\rm pert}(\eta_*)]
+O(\epsilon^4,\delta^{n+1}).
\label{eq: matching solution zeta}
\end{eqnarray} 
%
%
%%%%%%%%%%%%%%%%%%%%%%%%%%%%%%%%%%%%%%%
This is the general solution matched to ${\cal R}_c^{\rm pert}$ 
at $\eta=\eta_*$. 
What we need to know is the final value of ${\cal R}_c^{\rm NL}$
at sufficiently late times, $\eta\to0$ ($t\to\infty$).
It is 
\begin{eqnarray}
{\cal R}_c^{\rm NL}(0)={\zeta}_{c}(0)
={\cal R}_c^{\rm pert}(\eta_*)+
{D_*\over 3 {\cal H}_*}({{\cal R}_c^{\rm pert}})'(\eta_*)-
{F_*\over 4}K^{(2)}[{\cal R}_c^{\rm pert}(\eta_*)]+O(\epsilon^4,\delta^{n+1}),
\label{eq: zeta-infty}
\end{eqnarray}
where note the first equality which follows from
the assumption (\ref{eq: gamma-ij t-infty}): $E_c\to 0$
at sufficiently late times. Parallel to the second-order differential
equation (\ref{eq: basic eq for NL}) for ${\cal R}_c^{\rm NL}$,
which is a natural extension of the well-known linear 
version~(\ref{eq: linear eq R}), the above expression for the final
value of ${\cal R}_c^{\rm NL}$ turns out to be a natural extension
of the result obtained in linear theory in \cite{Leach:2001zf}.

%%%%%%%%%%%%%%%%%%%%%%%%%%%%%%%%%%%%%
%
%
\subsection{$t_*$-independence}
%
%
%%%%%%%%%%%%%%%%%%%%%%%%%%%%%%%%%%%%%%

In Sec.~\ref{subsec:tkshift}, we considered the variation of 
${\cal R}_c^{\rm NL}$ under an infinitesimal shift of $t_*$,
and derived consistency conditions on the variation of the 
undetermined spatial functions 
${\tilde\ell}^{(2)}$ and ${\tilde C}^{(2)}$.
Here we show that the final result (\ref{eq: zeta-infty})
obtained by the matching at $t=t_*$ is indeed $t_*$-independent.

If we take the variation of ${\cal R}_c^{\rm NL}(0)$ given by
(\ref{eq: zeta-infty}) with respect to the matching
time $t_*$, we obtain up to errors of $O(\epsilon^4,\delta^{n+1})$,
\begin{eqnarray}
\delta_{t_*} {\cal R}_c^{\rm NL}(0)
&=&\delta\eta\left[({\cal R}_c^{\rm pert})'+
{D_*\over 3 {\cal H}_*}({{\cal R}_c^{\rm pert}})''
+\left({D_*\over 3 {\cal H}_*}\right)'({{\cal R}_c^{\rm pert}})'
-{F_*'\over 4}K^{(2)}[{\cal R}_c^{\rm pert}]\right](\eta_*)
\nonumber\\
&=&\delta\eta\left[({{\cal R}_c^{\rm pert}})'+
{D_*\over 3 {\cal H}_*} ({{\cal R}_c^{\rm pert}})''
-({{\cal R}_c^{\rm pert}})'
\left(1-2{z'\over z}{D_*\over 3 {\cal H}_*}\right)+{D_*\over 12 {\cal H}_*}
 c_s^2 K^{(2)}[{{\cal R}_c^{\rm pert}}] \right](\eta_*)
\nonumber\\
&=&{D_*\delta\eta\over 3 {\cal H}_*}\left[({{\cal R}_c^{\rm pert}})''
+2{z'\over z}({{\cal R}_c^{\rm pert}})'
+\frac{c_s^2}{4} K^{(2)}[{{\cal R}_c^{\rm pert}}] \right](\eta_*)\,,
\end{eqnarray}
where $\delta\eta=\delta t/a_*$. Thus for ${\cal R}_c^{\rm NL}(0)$
to be $t_*$-independent, ${\cal R}_c^{\rm pert}$ must satisfy
\begin{eqnarray}
({{\cal R}_c^{\rm pert}})''+2{z'\over z}
({{\cal R}_c^{\rm pert}})'
+{c_s^2\over 4} K^{(2)}[{\cal R}_c^{\rm pert}]
=O(\epsilon^4,\delta^{n+1})
\quad\mbox{at}~\eta=\eta_*\,.
\end{eqnarray}
This is exactly what we assumed for ${\cal R}_c^{\rm pert}$:
It should satisfy (\ref{eq: basic eq for NL}) except for additional
errors of $O(\delta^{n+1})$.
Hence we conclude that ${\cal R}_c^{\rm NL}(0)$ is indeed
$t_*$-independent.

%%%%%%%%%%%%%%%%%%%%%%%%%%%%%%%%%%%%%%%%%%%%%%%%%%%%%%%%%%%%%%%%%%%%%%%%%
%
%
%
\section{Matching linear solution to nonlinear solution}
\label{sec:linearmatch}
%
%
%%%%%%%%%%%%%%%%%%%%%%%%%%%%%%%%%%%%%%%%%%%%%%%%%%%%%%%%%%%%%%%%%%

In order to determine the nonlinear solution ${\cal R}_c^{\rm NL}$,
we need to know the values of the $n$-th-order perturbative solution
${\cal R}_c^{\rm pert}$ and its first time derivative at the matching
time $\eta=\eta_*$. In general this is a formidable task. However, if we  
consider the case in which the universe is in the conventional,
slow-roll single-field inflation at the stage $\eta<\eta_*$,
the linear solution is a very good approximation on both 
subhorizon and superhorizon scales.
Even if intrinsic nonlinearity is important for subhorizon scale 
quantum fluctuations, such as the case of DBI inflation,
there may be cases in which the evolution near the horizon crossing time
may be well approximated by linear theory. 
In this section, we focus on such a case, that is, 
the case in which the nonlinear solution ${\cal R}_c^{\rm NL}$ and its 
first time derivative can be determined with sufficient accuracy by the
linear solution ${\cal R}^{\rm Lin}_c$ at the horizon crossing time.

Note that, while there is no problem in defining Fourier components 
${\cal R}_{c,\bm{k}}^{\rm NL}$ of  the nonlinear curvature perturbation
and thus the corresponding horizon crossing time $\eta=\eta_k$, Fourier
components with different $\bm{k}$ do not evolve independently. Hence,
we should use the same matching time for all Fourier
components. Otherwise, it would not be obvious whether matching
conditions for different Fourier components are consistent with each
other. This is the reason why we have introduced $\eta_*$.

Thus the nonlinear solution should be obtained by the replacements,
\begin{eqnarray}
{\cal R}_c^{\rm pert}(\eta_*)
 \to {\cal R}_c^{\rm Lin}(\eta_*) + s_1(\eta_*)\,,
\quad
{{\cal R}_c^{\rm pert}}'(\eta_*)
\to {{\cal R}_c^{\rm Lin}}'(\eta_*) + s_2(\eta_*)\,,
\end{eqnarray}
in the right-hand side of (\ref{eq: matching solution zeta}), where
$s_{1,2}(\eta_*)=O(\delta^2)$ are functions of $\eta_*$ and spatial
coordinates. This boundary condition uniquely determines 
${\cal R}_c^{\rm NL}$ and thus its Fourier components 
${\cal R}_{c,\bm{k}}^{\rm NL}(\eta)$, provided 
that $s_{1,2}(\eta_*)$ are specified. The nonlinear part
$s_{1,2}(\eta_*)$ of the matching condition is determined by requiring 
that the resulting ${\cal R}_{c,\bm{k}}^{\rm NL}(\eta)$ and its time
derivative ${{\cal R}_{c,\bm{k}}^{\rm NL}}'(\eta)$ evolved backward 
in time do not include nonlinear terms at $\eta=\eta_k$. This
requirement is nothing but a restatement of our assumption that 
the evolution near the horizon crossing time be well approximated by
linear theory.

Below we first briefly review the general linear solution on
superhorizon scales obtained by Leach et al.~\cite{Leach:2001zf}.
Then we spell out the nonlinear solution in terms of the linear
solution. Finally, we derive the bispectrum from our solution
assuming that the linear solution is a Gaussian random field.

%%%%%%%%%%%%%%%%%%%%%%%%%%%%%%%%%%%%%%%%%%%%%%%%%%%%%
%
%
\subsection{Linear solution valid up to $O(\epsilon^2)$}
%
%
%%%%%%%%%%%%%%%%%%%%%%%%%%%%%%%%%%%%%%%%%%%%%%%%%%%%%

In linear theory, the curvature perturbation on comoving hypersurfaces
follows (\ref{eq: linear eq R}). As usual, we consider it in the
Fourier space,
\begin{eqnarray}
{{\cal R}^{\rm Lin}_{c,\bm{k}}}''+2{z'\over z} {{\cal R}^{\rm Lin}_{c,\bm{k}}}'
+c_s^2k^2\,{\cal R}^{\rm Lin}_{c,\bm{k}}=0\,.
\label{eq: linearFeq R}
\end{eqnarray}
Real space expressions will be recovered by the replacement $k^2\to-\Delta$
at the end of calculation.

The above equation has two independent solutions; conventionally 
called a growing mode and a decaying mode. We assume that the growing mode
is constant in time at leading order in the long-wavelength approximation
or equivalently in the spatial gradient expansion. 
Then in terms of the growing mode solution, $u$, the 
decaying mode solution, $v$, can be given as \cite{Leach:2001zf}
\begin{eqnarray}
v(\eta)=u(\eta){\tilde{D}(\eta)\over \tilde{D}(\eta_*)}\,;
\quad
\tilde{D}(\eta)=3 {\cal H}_* \int_\eta^{0}
d\eta'{z^2(\eta_*)u^2(\eta_*)\over z^2(\eta')u^2(\eta')}\,.
\label{def: v-linear}
\end{eqnarray}
The general solution of a curvature perturbation is 
written in terms of their linear combinations as 
\begin{eqnarray}
{\cal R}^{\rm Lin}_c(\eta)
 =\alpha^{\rm Lin} u(\eta)+\beta^{\rm Lin} v(\eta)\,,
\label{eq: linear sol-R0}
\end{eqnarray}
where the coefficients $\alpha^{\rm Lin}$ and $\beta^{\rm Lin}$ 
may be assumed to satisfy $\alpha^{\rm Lin}+\beta^{\rm Lin}=1$
without loss of generality. Note that the assumption of the 
gradient expansion (\ref{eqn:assumption-gamma}) corresponds to
the condition,
\begin{eqnarray}
\beta^{\rm Lin}_{\bm{k}}=1-\alpha^{\rm Lin}_{\bm{k}}=O(k^2)\,.
\label{eq: assumption-alpha}
\end{eqnarray}
This means, as mentioned before, that the decaying mode at
leading order in the gradient expansion has already decayed
after horizon crossing.

First we solve for the growing mode solution. 
In accordance with the gradient expansion, we set
\begin{eqnarray}
u_{\bm{k}}(\eta)=\sum^\infty_{n=0} u_n(\eta) k^{2 n}\,.
\end{eqnarray}
At leading order in the gradient expansion, 
the growing mode solution $u^{(0)}_{\bm{k}}$ is just a constant.
Then inserting the above expansion with $u^{(0)}_{\bm{k}}=$const.
to the equation of motion~(\ref{eq: linearFeq R}) gives
iteratively
\begin{eqnarray}
u''_{n+1}+2{z'\over z} u'_{n+1}=-c_s^2u_{n}\,.
\end{eqnarray}
As shown in \cite{Leach:2001zf}, 
$O(k^2)$ corrections to $u^{(0)}_{\bm{k}}$ 
can be written as 
\begin{eqnarray}
u^{(2)}_{\bm{k}}=u^{(0)}_{\bm{k}}\left[C_1^{(2)} +C_2^{(2)} 
D(\eta)+k^2F(\eta)
\right],
\end{eqnarray}
where the integrals $D(\eta)$ and $F(\eta)$ have been given 
in (\ref{def: integral-D-F}), and $C_1^{(2)}$ and $C_2^{(2)}$
are arbitrary constants. 
We fix the two arbitrary constants as
$C_1^{(2)}=0$ and $C_2^{(2)}=-k^2F_*/D_*$ so that
$u_{\bm{k}}(\eta_*)=u^{(0)}_{\bm{k}}$ holds at $O(k^2)$ accuracy.
Hence 
\begin{eqnarray}
u^{(2)}_{\bm{k}}(\eta)
=\left[-{F_* \over D_*}D(\eta)+F(\eta)\right]k^2u^{(0)}_{\bm{k}}\,.
\label{sol: Leach Sasaki}
\end{eqnarray}

As for the decaying mode, since the coefficient $\beta^{\rm Lin}_{\bm{k}}$ is
already of $O(k^2)$, we only need the leading order solution.
Since we may replace $\tilde D$ with $D$ in (\ref{def: v-linear}),
we immediately find
\begin{eqnarray}
v^{(0)}_{\bm{k}}(\eta)=u^{(0)}_{\bm{k}}{D(\eta)\over D_*}\,.
\label{eq: linear sol-R}
\end{eqnarray}
Thus from (\ref{sol: Leach Sasaki}) and (\ref{eq: linear sol-R}), 
the general linear solution valid up to $O(\epsilon^2)$ is obtained
as
\begin{eqnarray}
{\cal R}_{c,\bm{k}}^{\rm Lin}(\eta)
=\left[\alpha^{\rm Lin}_{\bm{k}}+(1-\alpha^{\rm Lin}_{\bm{k}}
){D(\eta)\over D_*}
+\left(-{F_* \over D_*}D(\eta)+F(\eta)\right)k^2\right]u^{(0)}_{\bm{k}}\,.
\label{sol: linear Rc}
\end{eqnarray}
Note that ${\cal R}^{\rm Lin}_{c,_{\bm{k}}}(\eta_*)=u^{(0)}_{\bm{k}}$
while ${\cal R}^{\rm Lin}_{c,_{\bm{k}}}(0)=\alpha^{\rm Lin}_{\bm{k}}u^{(0)}
_{\bm{k}}$.
Thus if the factor $|\alpha^{\rm Lin}_{\bm{k}}|$ is large, it represents 
an enhancement of the curvature perturbation on superhorizon scales
due the $O(\epsilon^2)$ effect, which happens when
the slow-roll conditions are violated.
This will be discussed in detail in Sec.~VI. 

%%%%%%%%%%%%%%%%%%%%%%%%%%%%%%%%%%%%%%%%%%%%%%%%%%%%%%

Here it is useful to consider an explicit
expression for $\alpha^{\rm Lin}_{\bm{k}}$ in terms of 
${\cal R}^{\rm Lin}_{c,{\bm{k}}}$ and its derivative at $\eta=\eta_*$.
 From the general solution given by (\ref{eq: linear sol-R0}), we have
\begin{eqnarray}
{\cal R}^{\rm Lin}_{c,{\bm{k}}}(\eta)
 &=&\alpha^{\rm Lin}_{\bm{k}} u_{\bm{k}}(\eta)
+(1-\alpha^{\rm Lin}_{\bm{k}}) v_{\bm{k}}(\eta)\,,
\nonumber\\
{{\cal R}^{\rm Lin}_{c,{\bm{k}}} }'(\eta)
 &=&\alpha^{\rm Lin}_{\bm{k}} u'_{\bm{k}}(\eta)+
(1-\alpha^{\rm Lin}_{\bm{k}}) v'_{\bm{k}}(\eta)\,,
\end{eqnarray}
where we replaced $\beta^{\rm Lin}_{\bm{k}}$ by $1-\alpha^{Lin}_{\bm{k}}$.
Using the definition (\ref{def: v-linear}) of $v_{\bm{k}}(\eta)$,
the above equations evaluated at $\eta=\eta_*$ read
\begin{eqnarray}
&&{\cal R}^{\rm Lin}_{c,_{\bm{k}}}(\eta_*)=u_{\bm{k}}(\eta_*)\,,
\nonumber\\
&&{{\cal R}^{\rm Lin}_{c,_{\bm{k}}}}'(\eta_*)={u}_{\bm{k}}'(\eta_*)
-{3{\cal H}_* \over \tilde{D}_*}(1-\alpha^{\rm Lin}_{\bm{k}}) u_{\bm{k}}
(\eta_*)\,.
\end{eqnarray}
We may solve these equations for $\alpha^{\rm Lin}_{\bm{k}}$.
The result is 
\begin{eqnarray}
\alpha^{\rm Lin}_{\bm{k}}=1+ {\tilde{D}_*\over 3 {\cal H}_*}
\left[\frac{{{\cal R}^{\rm Lin}_{c,_{\bm{k}}}}'}{{\cal R}^{\rm Lin}_{c,_{\bm{k}}}}
-{u_{\bm{k}}'\over u_{\bm{k}}}\right]_{\eta=\eta_*}.
\end{eqnarray}
At $O(k^2)$ accuracy, we have
\begin{eqnarray}
{u'_{\bm{k}}(\eta_*)\over u_{\bm{k}}(\eta_*)}
=\frac{{{u}^{(2)}_{\bm{k}}}'(\eta_*)}{u^{(0)}_{\bm{k}}(\eta_*)}+O(k^4)
=\frac{3 {\cal H}_* F_*}{D_*}k^2+O(k^4)\,,
\end{eqnarray}
and hence 
\begin{eqnarray}
\beta^{\rm Lin}_{\bm{k}}=1-\alpha^{\rm Lin}_{\bm{k}}
=-\frac{D_*}{3 {\cal H}_*}
\frac{{{\cal R}^{\rm Lin}_{c,_{\bm{k}}}}'}
{{\cal R}^{\rm Lin}_{c,_{\bm{k}}}}\bigg|_{\eta=\eta_*} 
+k^2 F_*+O(k^4)\,.
\label{eq: alpha-linear0}
\end{eqnarray}

In order to relate our calculation with the standard formula for
the curvature perturbation in linear theory, we introduce 
$\eta_k$ (or $t_k$) which denotes the time at which the comoving wavenumber
has crossed the Hubble horizon,
\begin{equation}
\eta_k=-{r\over k}\,;\quad 0<r\ll 1 \,.
\end{equation}
The power spectrum at the horizon crossing time is given by 
%============< EQUATION >==============%
%
\begin{equation}
 \langle {\cal R}^{\rm Lin}_{c,\bm{k}}(\eta_k)
  {\cal R}^{\rm Lin}_{c,\bm{k}'}(\eta_{k'})
  \rangle
  = (2\pi)^3P_{{\cal R}}^{(0)}(k)\delta^3(\bm{k}+\bm{k}'), 
  \quad
  P_{{\cal R}}^{(0)}(k) = 
  \left|{\cal R}^{\rm Lin}_{c,\bm{k}}(\eta_k)\right|^2\,.
  \label{eq: power spectrum for R}
\end{equation}
%======================================%

As shown in the Appendix~\ref{app:alphau}, we can show the final value
of the linear curvature perturbation as 
\begin{equation}
{\cal R}^{\rm Lin}_{c,\bm{k}}(0)=
 \alpha^{\rm Lin}_{\bm{k}}u^{(0)}_{\bm{k}} = 
  \tilde{\alpha}^{\rm Lin}_{\bm{k}}
  {\cal R}^{\rm Lin}_{c,\bm{k}}(\eta_k)
  + O(k^4),
  \label{eqn:alphau}
\end{equation}
%======================================%
where
%============< EQUATION >==============%
%
\begin{equation}
 \tilde{\alpha}^{\rm Lin}_{\bm{k}}
  =
  1 + \alpha^{\cal R}\tilde{D}_k- k^2 \tilde{F}_k 
  \label{til-alpha-lin}
\end{equation}
%======================================%
and 
\begin{eqnarray}
 \alpha^{\cal R} & = & 
  \frac{1}{3{\cal H}(\eta_k) }
  {{\cal R}'_c\over {\cal R}_c}\bigg|_{\eta=\eta_k},
  \nonumber\\
 \tilde{D}_k & = & 3{\cal H}(\eta_k) 
  \int_{\eta_k}^0d\eta' {z^2(\eta_k) \over  z^2 (\eta')},
  \nonumber\\
 \tilde{F}_k & = &
  \int_{\eta_k}^0\frac{d\eta'}{z^2(\eta')}
  \int_{\eta_k}^{\eta'}z^2(\eta'')c_s^2(\eta'')d\eta''.
\end{eqnarray}
This explicitly shows that 
$\alpha^{\rm Lin}_{\bm{k}}u^{(0)}_{\bm{k}}$ is independent of 
$\eta_*$ up through $O(k^2)$. The formula (\ref{eqn:alphau}) will 
be used in the next subsection.

The power spectrum at the final time is thus enhanced by the factor 
$|\tilde{\alpha}^{\rm Lin}_{\bm{k}}|^2$ as
%============< EQUATION >==============%
%
\begin{equation}
 \langle {\cal R}^{\rm Lin}_{c,\bm{k}}(0)
  {\cal R}^{\rm Lin}_{c,\bm{k}''}(0)
  \rangle
  = (2\pi)^3  |\tilde{\alpha}^{\rm Lin}_{\bm{k}}|^2
  P_{{\cal R}}^{(0)}(k)\delta^3(\bm{k}+\bm{k}'). 
  \label{eqn:powerspectrum}
\end{equation}
%======================================%

%%%%%%%%%%%%%%%%%%%%%%%%%%%%%%%%%%%%%%%%%%%%%%%%%%%%%%%%
%
%    Matched nonlinear solution
%
%%%%%%%%%%%%%%%%%%%%%%%%%%%%%%%%%%%%%%%%%%%%%%%%%%
\subsection{Matched nonlinear solution}

Using the linear solution of the curvature perturbation 
given by (\ref{sol: linear Rc}), here we derive
the nonlinear solution by matching the two at $\eta=\eta_*$. 
The main purpose of the matching is to make it possible to analyze
super-horizon nonlinear evolution valid up to the second-order in
gradient expansion, starting from a solution in the linear theory. In
particular, we would like to evaluate the bispectrum induced by the
super-horizon nonlinear evolution. For this purpose, we need to
have full control over terms up not only to $O(\epsilon^2)$ but also
to $O(\delta^2)$, where we suppose that the linear solution is of order
$O(\delta)$. Therefore, the matching condition at $\eta=\eta_*$ should
be of the form 
%============< EQUATION >==============%
%
\begin{eqnarray}
 {\cal R}^{\rm NL}_c(\eta_*)
  & = & {\cal R}^{\rm Lin}_c(\eta_*) + s_1(\eta_*)
  + O(\epsilon^4,\delta^3), \nonumber\\
 {{\cal R}^{\rm NL}_c}'(\eta_*)
  & = & {{\cal R}^{\rm Lin}_c}'(\eta_*)
  + s_2(\eta_*) + O(\epsilon^4,\delta^3),
\end{eqnarray}
%======================================%
where 
%============< EQUATION >==============%
%
\begin{equation}
 s_{1}(\eta_*) = O(\delta^2)\,,
\quad s_{2}(\eta_*) = O(\delta^2)
  \label{eqn:R2delta2}
\end{equation}
%======================================%
are functions of $\eta_*$ and spatial coordinates. While the linear
solution ${\cal R}^{\rm Lin}_c(\eta)$ is considered as an input,
i.e., initial condition, the additional terms, $s_{1}(\eta_*)$ and
$s_2(\eta_*)$, are to be determined by the following condition: 
\begin{itemize}
\item
     The terms of order $O(\delta^2)$ in 
     ${\cal R}^{\rm NL}_{c,\bm{k}}$ and 
     ${{\cal R}^{\rm NL}_{c,\bm{k}}}'$ should vanish at $\eta=\eta_k$, 
     where ${\cal R}^{\rm NL}_{c,\bm{k}}$ is the Fourier component of 
     ${\cal R}^{\rm NL}_c$ and $\eta_k$ is the time slightly after
     the wavenumber $k=|\bm{k}|$ has crossed the horizon; $-k\eta_k=r\ll1$.
\\
\end{itemize}
In other words, $s_1(\eta_*)$ and $s_2(\eta_*)$ represent the
$O(\delta^2)$ part of ${\cal R}^{\rm NL}_c$ and 
${{\cal R}^{\rm NL}_c}'$, respectively, generated during the period between
the horizon crossing time and the matching time.

The matching condition (\ref{eq: matching cond1}) and 
(\ref{eq: matching cond2}) is written explicitly as
%============< EQUATION >==============%
%
\begin{eqnarray}
 \ell^{(0)} + \ell^{(2)} +H^{(2)}  & = &
  {\cal R}^{\rm Lin}_c(\eta_*) + s_1(\eta_*)
  + O(\epsilon^4,\delta^3), \nonumber\\
 C^{(2)} & = & -\frac{1}{3{\cal H}_*}
  \left( {{\cal R}^{\rm Lin}_c}'(\eta_*)
   + s_2(\eta_*)
  \right)
  + O(\epsilon^4,\delta^3). 
\end{eqnarray}
%======================================%
The nonlinear solution (\ref{eq: matching solution zeta})
is now expressed as 
%============< EQUATION >==============%
%
\begin{eqnarray}
 {\cal R}^{\rm NL}_c(\eta) & = & 
  \left[{\cal R}^{\rm Lin}_c(\eta_*) + s_1(\eta_*) \right]
  + \frac{D_*-D(\eta)}{3{\cal H}_*}
  \left[{{\cal R}^{\rm Lin}_c}'(\eta_*) + s_2(\eta_*) \right]
  \nonumber\\
 & & 
  - \left[F(\eta)-F_*\right] \Delta
  \left[{\cal R}^{\rm Lin}_c(\eta_*) + s_1(\eta_*) \right]
  + \frac{1}{4}\left[F(\eta)-F_*\right]
  \tilde{K}^{(2)}[{\cal R}^{\rm Lin}_c(\eta_*)]
  + O(\epsilon^4, \delta^3),
\end{eqnarray}
%======================================%
where
%============< EQUATION >==============%
%
\begin{equation}
 \tilde{K}^{(2)}[\ell^{0}] \equiv 
  -2\left(\delta^{ij}\partial_i\ell^{0}\partial_j\ell^{0}-4
  \ell^{0}\Delta \ell^{0}\right)
  = 4\Delta \ell^{0} + K^{(2)}[\ell^{0}] + O((\ell^{0})^3).
\end{equation}
%======================================%
The corresponding Fourier component is
%============< EQUATION >==============%
%
\begin{eqnarray}
 {\cal R}^{\rm NL}_{c,\bm{k}}(\eta) & = & 
  \left\{
   {\cal R}^{\rm Lin}_{c,\bm{k}}(\eta_*) 
   + \frac{D_*-D(\eta)}{3{\cal H}_*}
   {{\cal R}^{\rm Lin}_{c,\bm{k}}}'(\eta_*)
   - \left[F_*-F(\eta)\right] k^2{\cal R}^{\rm Lin}_{c,\bm{k}}(\eta_*)
   \right\}
  \nonumber\\
 & +& 
  \left\{
   s_{1,\bm{k}}(\eta_*)
   + \frac{D_*-D(\eta)}{3{\cal H}_*} s_{2,\bm{k}}(\eta_*)
   - \left[F_*-F(\eta)\right] 
   \left[
    k^2 s_{1,\bm{k}}(\eta_*)
    + \frac{1}{4}
    \tilde{K}^{(2)}_{\bm{k}}[{\cal R}^{\rm Lin}_c(\eta_*)]
       \right]
    \right\}
  + O(\epsilon^4, \delta^3).
  \label{eqn:NLsol}
\end{eqnarray}
%======================================%
Note that, as already stated, the Fourier components with different
$\bm{k}$ do not evolve independently.

By demanding that the terms of order $O(\delta^2)$ in 
${\cal R}^{\rm NL}_{c,\bm{k}}$ and 
${{\cal R}^{\rm NL}_{c,\bm{k}}}'$ should vanish at $\eta=\eta_k$, 
$s_{1,\bm{k}}(\eta_*)$ and $s_{2,\bm{k}}(\eta_*)$ are determined as 
%============< EQUATION >==============%
%
\begin{eqnarray}
 s_{1,\bm{k}}(\eta_*) & = & 
  -\frac{1}{4}\tilde{K}^{(2)}_{\bm{k}}[{\cal R}^{\rm Lin}_c(\eta_*)]
  \int_{\eta_k}^{\eta_*}\frac{d\eta'}{z^2(\eta')}
  \int_{\eta_k}^{\eta'}z^2(\eta'')c_s^2(\eta'')d\eta'' 
  + O(\epsilon^4,\delta^3), \nonumber\\
 s_{2,\bm{k}}(\eta_*) & = & 
  -\frac{1}{4}
  \tilde{K}^{(2)}_{\bm{k}}[{\cal R}^{\rm Lin}_c(\eta_*)]
  \times
  \frac{1}{z^2(\eta_*)}
   \int_{\eta_k}^{\eta_*}z^2(\eta')c_s^2(\eta')d\eta' 
  + O(\epsilon^4,\delta^3),
\end{eqnarray}
%======================================%
Therefore, by substituting these to (\ref{eqn:NLsol}), we obtain 
%============< EQUATION >==============%
%
\begin{eqnarray}
 {\cal R}^{\rm NL}_{c,\bm{k}}(\eta)  &=&  
  {\cal R}^{\rm Lin}_{c,\bm{k}}(\eta_*)
  + \frac{D_*-D(\eta)}{3{\cal H}_*}
  {{\cal R}^{\rm Lin}_{c,\bm{k}}}'(\eta_*)
\cr
&&\quad
  + \left[F(\eta)-F_*\right] k^2
  {\cal R}^{\rm Lin}_{c,\bm{k}}(\eta_*)
  - \frac{1}{4}{\cal F}_k(\eta)
  \tilde{K}^{(2)}_{\bm{k}}[{\cal R}^{\rm Lin}_c(\eta_*)]
  + O(\epsilon^4, \delta^3),
  \label{eq: matching solution R-linear}
\end{eqnarray}
%======================================%
where
%============< EQUATION >==============%
%
\begin{equation}
 {\cal F}_k(\eta) = 
  \int_{\eta_k}^{\eta}\frac{d\eta'}{z^2(\eta')}
  \int_{\eta_k}^{\eta'}c_s^2(\eta'')z^2(\eta'')d\eta''. 
\end{equation}

Using the linear solution of the curvature perturbation 
given by (\ref{sol: linear Rc}), we have 
\begin{eqnarray}
{\cal R}_{c,\bm{k}}^{\rm Lin}(\eta_*)=u^{(0)}_{\bm{k}}\,,\quad
{{\cal R}^{\rm Lin}_{c,\bm{k}}}'(\eta_*)=-\frac{3 {\cal H}_*}{D_*}
(1-\alpha^{\rm Lin}_{\bm{k}} - k^2F_*) u^{(0)}_{\bm{k}}.
\end{eqnarray}
Substituting these into (\ref{eq: matching solution R-linear}), taking
the limit $\eta\to 0$ and using (\ref{eqn:alphau}) yield the nonlinear
comoving curvature perturbation at the final time $\eta=0$ (or
$t=\infty$) given by 
\begin{eqnarray}
{\cal R}_{c,\bm{k}}^{\rm NL}(0)&=&
\tilde{\alpha}^{\rm Lin}_{\bm{k}}
{\cal R}^{\rm Lin}_{c,\bm{k}} (\eta_k) 
-\frac{1}{4}{\cal F}_k(0)
  \tilde{K}^{(2)}_{\bm{k}}[{\cal R}^{\rm Lin}_c(\eta_k)]
  +O(\epsilon^4, \delta^3)
\nonumber\\
&=&{\cal R}^{\rm Lin}_{c,\bm{k}} (\eta_k) 
-(1-\tilde{\alpha}^{\rm Lin}_{\bm{k}})
{\cal R}^{\rm Lin}_{c,\bm{k}} (\eta_k) 
-\frac{1}{4}{\cal F}_k(0)
  \tilde{K}^{(2)}_{\bm{k}}[{\cal R}^{\rm Lin}_c(\eta_k)]
 +O(\epsilon^4, \delta^3)\,.
\label{sol: infty Rreal}
\end{eqnarray}
This is the main result. 
The first term corresponds to the result of the $\delta N$ formalism,
the second term is related to an enhancement on superhorizon scales
in linear theory, and the last term is the nonlinear effect which
may become important if ${\cal F}_k(0)$ is large.

%%%%%%%%%%%%%%%%%%%%%%%%%%%%%%%%%%%%%%%%%%%%%%%%%%%%%%%%%%%%%
\subsection{Bispectrum}
In this subsection, we calculate the bispectrum of our nonlinear 
curvature perturbation by assuming that 
${\cal R}^{\rm Lin}_{c,\bm{k}}(\eta_k)$ is a Gaussian 
random variable. We assume the leading order contribution to the bispectrum
comes from the terms second order in ${\cal R}^{\rm Lin}_{c,{\bm{k}}}(\eta_k)$. 
The final result (\ref{sol: infty Rreal}) gives
\begin{eqnarray}
 \zeta_{\bm{k}}&=&{\cal R}^{\rm NL}_{c,\bm{k}}(0) \nonumber\\
 & = & 
  G(\bm{k})\ {\cal R}^{\rm Lin}_{c,\bm{k}}(\eta_k)
 + H(k)\ 
 \left\{
  \int\frac{d^3k'd^3k''}{(2\pi)^3}
    (4{k'}^2-\delta_{ij}{k'}^i{k''}^j)
    {\cal R}^{\rm Lin}_{c,\bm{k}'}(\eta_{k'})
    {\cal R}^{\rm Lin}_{c,\bm{k}''}(\eta_{k''})
    \delta^3(-\bm{k}+\bm{k}'+\bm{k}'')
 \right\} \nonumber\\
 & & 
  + O(\epsilon^4, \delta^3),
  \label{eq: final-zeta-k}
\end{eqnarray}
%======================================%
where
%============< EQUATION >==============%
%
\begin{eqnarray}
 G(\bm{k}) & \equiv & \tilde{\alpha}^{\rm Lin}_{\bm{k}}
  = 1 + \alpha^{\cal R}\tilde{D}_k - k^2\tilde{F}_k, \nonumber\\
  H(k) & \equiv & \frac{1}{2}{\cal F}_k(0) = \frac{1}{2}\tilde{F}_k.
  \label{eqn:def-GH}
\end{eqnarray}
%======================================%
This is independent of $\eta_*$ as should be. Note also that, in general 
$G(\bm{k})$ may depend on the directions of ${\bm k}$. However, in the
present case we may assume the absence of such spatial anisotropy:
$G(\bm{k})=G(k)$.

By assuming the Gaussian statistics for 
${\cal R}^{\rm Lin}_{c,\bm{k}}(\eta_k)$ with 
the power spectrum (\ref{eq: power spectrum for R}), 
it is easy to calculate the power spectrum and the bispectrum of
$\zeta$. 

The bispectrum $B_{\zeta}$ is expressed in terms of the 
Fourier transformation of the three point function as
\begin{eqnarray}
\left\langle{\zeta}_{{\bm k}_1}{\zeta}_{{\bm k}_2}{\zeta}_{{\bm k}_3}
\right\rangle_C
=(2\pi)^3B_{\zeta}({\bf k}_1, {\bf k}_2, {\bf k}_3)
\,\delta^{3} ({\bf k}_1+{\bf k}_2+{\bf k}_3)\,,
\end{eqnarray}
where $\langle\cdots\rangle_C$ means that it extracts out
only connected graphs.
With the help of (\ref{eq: final-zeta-k}), 
the three point correlation function of ${\zeta}$ is at leading order
calculated as
\begin{eqnarray}
&&\left\langle{\zeta}_{{\bm k}_1}{\zeta}_{{\bm k}_2}{\zeta}_{{\bm k}_3}
\right\rangle_C=(2\pi^3)\times
\nonumber\\
&&
\quad\left[G^*(k_1)G(k_2)H(k_3)
\left\{ 4(k_1^2+k_2^2)-2\delta_{ij}k_1^ik_2^j\right\}\delta^{(3)}
({\bf k}_1+{\bf k}_2+{\bf k}_3)
|{\cal R}^{\rm Lin}_{c,\bm{k}_1}(\eta_{k_1})|^2
|{\cal R}^{\rm Lin}_{c,\bm{k}_2}(\eta_{k_2})|^2+ {\rm perms} \right],
\end{eqnarray}
where a superscript star denotes a complex conjugate and `perms' means 
terms with permutations among the three wavenumbers.
The power spectrum of ${\cal R}^{\rm Lin}_{c,\bm{k}}(\eta_k)$ is written
as (\ref{eq: power spectrum for R}). Then we have 
\begin{eqnarray}
B_{\zeta}(k_1,k_2,k_3) & = & 
{G^*(k_1)\,G(k_2)\,H(k_3)}
\left\{5(k_1^2+k_2^2)-k_3^2\right\}
 P^{(0)}_{{\cal R}}(k_1)
 P^{(0)}_{{\cal R}}(k_2)
+ {\rm perms}
\nonumber\\
 & = & 
\frac{4\pi^4}{k_1^3k_2^3k_3^3}\left[
{G^*(k_1)\,G(k_2)\,H(k_3)}
\left\{5(k_1^2+k_2^2)-k_3^2\right\}
k_3^3\,
{\cal P}^{(0)}_{{\cal R}}(k_1)
{\cal P}^{(0)}_{{\cal R}}(k_2)
+ {\rm perms} \right]
\, ,
\label{eq: bispectrum-zeta}
\end{eqnarray}
where
\begin{equation}
{\cal P}^{(0)}_{\cal R}(k) = 
{k^3\over 2\pi^2}P^{(0)}_{\cal R}(k)
=
{k^3\over 2\pi^2}
\left|{\cal R}^{\rm Lin}_{c,\bm{k}}(\eta_k)\right|^2\,.
\end{equation}
We define the $k$-dependent $f_{NL}$ as
\begin{eqnarray}
B_{\zeta}(k_1,k_2,k_3)
 & = & \frac{6}{5}
 f_{NL}(k_1,k_2,k_3)
\left[
|\tilde{\alpha}^{\rm Lin}_{\bm{k}_1}
\tilde{\alpha}^{\rm Lin}_{\bm{k}_2}|^2
 P^{(0)}_{{\cal R}}(k_1)
 P^{(0)}_{{\cal R}}(k_2)
+ {\rm perms}
\right]\nonumber\\
 & = & \frac{24\pi^4}{5k_1^3k_2^3k_3^3}
 f_{NL}(k_1,k_2,k_3)
\left[
|\tilde{\alpha}^{\rm Lin}_{\bm{k}_1}
\tilde{\alpha}^{\rm Lin}_{\bm{k}_2}|^2
 {\cal P}^{(0)}_{{\cal R}}(k_1)
 {\cal P}^{(0)}_{{\cal R}}(k_2)
 k_3^3
+ {\rm perms}
\right]\, .
\end{eqnarray}
If ${\cal P}^{(0)}_{{\cal R}}(k)$ does not depend on $k$ then 
\begin{eqnarray}
f_{NL}(k_1,k_2,k_3)
 & = & \frac{5}{6}
\left[\sum_{i\neq j,j\neq k,k\neq i} 
|G(k_i)\,G(k_j)|^2\,k_k^3\right]^{-1}
\nonumber\\
& & \times
\left[\sum_{i\neq j,
j\neq k,k\neq i} G^*(k_i)\,G(k_j)\,H(k_k)\Bigl\{
5(k_i^2+ k_j^2)-k_k^2\Bigr\}k_k^3\right].
\label{eq: f-NL}
\end{eqnarray}
%%%%%%%%%%%%%%%%%%%%%%%%%%%%%%%

\section{Application to Starobinsky model}
\label{sec:starobinsky}

There are several known models in which the $O(\epsilon^2)$ effect
is important. For example, a potential of the form,
\begin{eqnarray}
V={\lambda\over 4} M^4 \left[1+B{64 \pi^2 \over m_{\rm pl}^4}
\phi^4\right],
\end{eqnarray} 
can lead to two separate stages of inflation with a temporary
suspension of slow-roll inflation in between the 
two stages~\cite{Leach:2000yw,Leach:2001zf}.
The Coleman-Weinberg potential can also lead to the same feature
as discussed in \cite{Saito:2008em}.
A similar feature is found in a theory with a
non-canonical Lagrangian~\cite{Jain:2007au},
\begin{eqnarray}
P=-V_0(1+V_1 \phi^4)\sqrt{1-X}\,,
\quad\mbox{with}~X=-g^{\mu\nu}\partial_\mu\phi\partial_\nu\phi\,.
\end{eqnarray}

In this section, we consider a simple model with a temporary
non-slow-roll stage first discussed by
Starobinsky \cite{Starobinsky:1992ts},
\begin{eqnarray}
V(\phi)=\left\{
\begin{array}{l}
V_0+A_{+}(\phi-\phi_0)\quad\mbox{for}~\phi>\phi_0\,,
\\
V_0+A_{-}(\phi-\phi_0)\quad\mbox{for}~\phi<\phi_0\,,
\end{array}
\right.
\label{def: Starobinsky model}
\end{eqnarray}
where $A_+>A_->0$ is assumed.
The advantage of this model is that it allows analytical
treatment of linear perturbations as well as of the background
evolution, provided that $V_0$ dominates in the potential,
$|V-V_0|/V_0\ll 1$.  If $A_+\gg A_-$,
and for $\phi$ initially large and positive,
the slow-roll condition is violated 
right after $\phi$ falls below $\phi_0$.
To be specific, the field enters a transient stage
at which $\ddot{\phi}\approx - 3H\dot{\phi}$ until the 
slow-roll condition is recovered: 
\begin{eqnarray}
3 H_0 \dot{\phi}=\left\{
\begin{array}{ll}
-A_+ &{\rm for~}\phi>\phi_0\,,\\
-A_- -(A_+ - A_-) e^{-3H_0 \Delta t}& {\rm for~}\phi<\phi_0\,,
\end{array}
\right.
%\label{def: Starobinsky model}
\end{eqnarray}
where $\Delta t=t-t_0$ with $t_0$ being the time at
which $\phi=\phi_0$ and the Hubble parameter $H$ is
approximated by $H_0=H(t_0)$ because $|V-V_0|\ll V_0$.
Then for $\Delta t>0$ we have
\begin{eqnarray}
\left|\frac{\ddot\phi}{3H_0\dot\phi}\right|
=\frac{A_+-A_-}{A_{-}e^{3H_0\Delta t}+A_+-A_-}\,.
\end{eqnarray}
Thus the slow-roll condition is violated during the stage
$3H_0\Delta t\lesssim \ln[(A_+-A_-)/A_-]$
if $A_+/A_-\gg 1$.

%%%%%%%%%%%%%%%%%%%%%%%%%%%%%%%%%%
\subsection{Linear solution}

For Starobinsky's model, $z$ defined by (\ref{def: variable-z}) 
is given by
\begin{eqnarray}
z=\frac{a_0}{3H_0^2}\times\left\{
\begin{array}{ll}
A_{+}e^{H_0\Delta t}&\quad\mbox{for}~ \Delta t<0\,,
\\
~\\
A_- e^{H_0 \Delta t}+(A_+-A_-)e^{-2H_0\Delta t}&
\quad\mbox{for}~\Delta t>0\,.
\end{array}\right.
\label{sol: starobinsky-z}
\end{eqnarray}
Substituting this into the integrals seen in 
(\ref{til-alpha-lin}), we obtain 
\begin{eqnarray}
\tilde{D}_k = 
\left\{
\begin{array}{ll}
\displaystyle 
1 + T\left(\frac{a_k}{a_0}\right)^3 &
\quad\mbox{for} \ \eta_k<\eta_0\ (t_k<t_0)\,,
\\
\displaystyle 
1 + T\left(\frac{a_0}{a_k}\right)^3 &
\quad\mbox{for} \ \eta_k>\eta_0\ (t_k>t_0)\,,
\end{array}
\right.
\label{eq: result-integrals-Dk}
\end{eqnarray}
and 
\begin{eqnarray}
\tilde{F}_k=\left\{
\begin{array}{ll}
\displaystyle
\frac{1}{k_0^2}
  \left(\frac{1}{6}\left(\frac{a_0}{a_k}\right)^2
   + \frac{2}{5}T
   - \frac{1}{3}T\frac{a_k}{a_0}\right) &
\quad\mbox{for} \ \eta_k<\eta_0\ (t_k<t_0)\,,
\\
~\\
\displaystyle
\frac{1}{k_0^2}
  \left(\frac{1}{6}\left(\frac{a_0}{a_k}\right)^2
   + \frac{1}{15}T\left(\frac{a_0}{a_k}\right)^5\right) &
\quad\mbox{for} \ \eta_k>\eta_0\ (t_k>t_0)\,,
\end{array}
\right.
\label{eq: result-integrals-Fk}
\end{eqnarray}
where $a_k=1/(-H_0\eta_k)=k/(rH_0)$, $k_0=a_0H_0$ is the comoving
wavenumber that crosses the horizon at $t=t_0$, and we have introduced
\begin{eqnarray}
T=\left({A_+\over A_-}-1\right)\,.
\end{eqnarray}
Thus if $A_+/A_-\gg 1$, we have $\tilde{D}_k\gg 1$ and
$k_0^2\tilde{F}_k\gg 1$ for wavenumbers $k$ not too much different from 
$k_0$. 

The perturbations with these wavenumbers must have left the horizon
before the transition time $\eta=\eta_0$. 
Then the vacuum mode function for the comoving curvature perturbation 
of Starobinsky's model before the transition is given by the standard
formula for slow-roll inflation,
\begin{eqnarray}
{\cal R}^{\rm Lin}_{c,\bm{k}}(\eta)
 =-{iH_0^2 \over \sqrt{2 k^3} \dot{\phi}}e^{- i k\eta}
(1+i k \eta)\,,
\end{eqnarray}
where $\dot\phi=-A_+/(3H_0)$. This gives 
\begin{eqnarray}
{({\cal R}^{\rm Lin}_{c,\bm{k}})'(\eta)\over {\cal R}_{c,\bm{k}}^{\rm Lin}}
={k^2\eta \over (1+{ik\eta})}\,,
\end{eqnarray}
and hence
\begin{equation}
\alpha^{\cal R}
=-\frac{k^2 \eta_k^2 }{3(1+ik\eta_k)}
=-{r^2\over 3(1-ir)}
\end{equation}
for $\eta_k< \eta_0$. 

The amplification factor $\tilde{\alpha}^{\rm Lin}_{\bm{k}}$ is given by
(\ref{til-alpha-lin}) with $\alpha^{\cal R}$, $\tilde{D}_k$ and
$\tilde{F}_k$ calculated above. Thus, for $\eta_k< \eta_0$, we have 
%============< EQUATION >==============%
%
\begin{eqnarray}
 \tilde{\alpha}^{\rm Lin}_{\bm{k}}
& = &
  1-\frac{r^2}{3(1-ir)}\left(1+T\frac{k^3}{r^3k_0^3}\right)
  - \frac{k^2}{k_0^2}
  \left(\frac{r^2}{6}\frac{k_0^2}{k^2}+\frac{2}{5}T
   -\frac{T}{3}\frac{k}{rk_0}\right)
\cr
&=&1-\frac{2}{5}\frac{k^2}{k_0^2}T-\frac{i}{3(1-ir)}\frac{k^3}{k_0^3}T
-\frac{r^2}{6}\frac{3-ir}{1-ir}\,.
\end{eqnarray}
%======================================%
For $T\gg 1$ and $k/k_0< r\ll\min\bigl[1,(k/k_0)T^{1/3}\bigr]$,
the power spectrum of the asymptotic value of the curvature perturbation is 
%============< EQUATION >==============%
%
\begin{eqnarray}
 \langle {\cal R}^{\rm Lin}_{c,\bm{k}}(0)
  {\cal R}^{\rm Lin}_{c,\bm{k}''}(0)
  \rangle
  & = & (2\pi)^3  
  P_{{\cal R}}\delta^3(\bm{k}+\bm{k}'), \nonumber\\
  P_{{\cal R}} & =  &
  |\tilde{\alpha}^{\rm Lin}_{\bm{k}}
  {\cal R}^{\rm Lin}_{c,\bm{k}}(\eta_k)|^2 
 \approx 
 \frac{H^4}{2k^3\dot{\phi}^2}
 \left[
  \left( 1 - \frac{2}{5}\frac{k^2}{k_0^2}T\right)^2
  +\frac{1}{9}\frac{k^6}{k_0^6}T^2
	       \right].
 \label{eqn:amplifiedPR}
\end{eqnarray}
%======================================%
Note that the inequality $k/k_0<r$ is required by
$\eta_k\leq\eta_*<\eta_0$, i.e., the condition that the linear
solution be matched to the nonlinear solution no later than $\eta_0$. 
This expression for the power spectrum is known to agree with the
exact result very well even for $Tk^2/k_0^2\gg 1$~\cite{Leach:2001zf} 
as long as $\eta_k<\eta_0$. 

%%%%%%%%%%%%%%%%%%%%%%%%%%%%%%%%%%%%%%%%%%%%%%%%%%%%%%%%

\subsection{Nonlinear solution}

In this subsection, we match the linear solution of
Starobinsky's model to the nonlinear solution on superhorizon scales
by applying the formulas obtained in Sec.~\ref{sec:linearmatch}.
We focus on the wavenumbers $k<k_0$. 
Recall (\ref{eq: final-zeta-k}) which gives the final amplitude
of the comoving curvature perturbation,
\begin{eqnarray}
 \zeta_{\bm{k}} 
 & = & 
  G(\bm{k})\ {\cal R}^{\rm Lin}_{c,\bm{k}}(\eta_k)
 + H(k)\ 
 \left\{
  \int\frac{d^3k'd^3k''}{(3\pi)^3}
    (4{k'}^2-\delta_{ij}{k'}^i{k''}^j)
    {\cal R}^{\rm Lin}_{c,\bm{k}'}(\eta_{k'})
    {\cal R}^{\rm Lin}_{c,\bm{k}''}(\eta_{k''})
    \delta^3(-\bm{k}+\bm{k}'+\bm{k}'')
 \right\} \nonumber\\
 & & 
  + O(\epsilon^4, \delta^3),
\end{eqnarray}
where $G(\bm{k})$ and $H(k)$ are defined in (\ref{eqn:def-GH}). 
In the present case they are given by 
\begin{eqnarray}
&&G(k)=
1-\frac{2}{5}\frac{k^2}{k_0^2}T-\frac{i}{3(1-ir)}\frac{k^3}{k_0^3}T
-\frac{r^2}{6}\frac{3-ir}{1-ir}\,,
\nonumber\\
&&H(k)=\frac{1}{5 k^2}\left[\frac{k^2}{k_0^2}T
\left(1-\frac{5}{6}\frac{k}{rk_0}\right)
    +\frac{5}{12}r^2\right]\,.
\label{eq: prefinal-result-G-H}
\end{eqnarray}

As mentioned at the end of the previous subsection,
the power spectrum given by ignoring the $r$-dependent terms
in $G(k)$ agrees well with the exact result without using
the long wavelength approximation. This strongly indicates that
the $r$-dependence of $G(k)$ is an artifact due to incomplete
matching of the exact linear solution with an approximate longwavelength
solution. This observation is supported by the fact that it 
disappears in the limit $r\to0$. Similarly, the $r$-dependence of
$H(k)$ must be also an artifact due to incomplete matching
of the linear and nonlinear solutions. In this case, however, 
the term $k/(rk_0)$ does not disappear in the limit $r\to0$.
This is because there should be a sufficient lapse of time
for the nonlinear solution to evolve before the transition time
to erase the memory of small errors in the initial condition,
implying that the accuracy of the nonlinear solution increases
on larger scales $k/k_0\ll r$. In short, we should take
the limit $k/k_0\ll r\ll\min\bigl[1,(k/k_0)T^{1/3}\bigr]$
in (\ref{eq: prefinal-result-G-H}) to get rid of the
$r$-dependence to obtain
\begin{eqnarray}
G(k)\approx
1-\frac{2}{5}\frac{k^2}{k_0^2}T-\frac{i}{3}\frac{k^3}{k_0^3}T\,,
\quad
H(k)\approx\frac{1}{5 k_0^2}T\,.
\label{eq: final-result-G-H}
\end{eqnarray}

%%%%%%%%%%%%%%%%%%%%%%%%%%%%%%%%%%%%%%%

\subsection{Bispectrum}

Finally, we estimate the bispectrum $B_\zeta$ 
and the corresponding non-Gaussianity parameter
 $f_{NL}({k}_1, {k}_2, { k}_3)$ given by 
 (\ref{eq: bispectrum-zeta}) and (\ref{eq: f-NL}), respectively. 
With the help of (\ref{eq: final-result-G-H}), we obtain
\begin{eqnarray}
B_{\zeta}(k_1,k_2,k_3) & = &
\frac{4\pi^4({\cal P}^{(0)}_{{\cal R}})^2}{k_1^3k_2^3k_3^3}
\sum_{i\neq j,j\neq k,k\neq i} 
{G^*(k_i)\,G(k_j)\,H(k_k)}
\left\{5(k_i^2+k_j^2)-k_k^2\right\}
k_k^3\, ,
\nonumber\\
f_{NL}(k_1,k_2,k_3)
 & = & \frac{5}{6}
\left[\sum_{i\neq j,j\neq k,k\neq i} 
|G(k_i)\,G(k_j)|^2\,k_k^3\right]^{-1}
\nonumber\\
& & \times
\left[\sum_{i\neq j,
j\neq k,k\neq i} G^*(k_i)\,G(k_j)\,H(k_k)\Bigl\{
5(k_i^2+ k_j^2)-k_k^2\Bigr\}k_k^3\right],
\label{eq: f-NL1}
\end{eqnarray}
where, for $T\gg 1$ and $k/k_0\ll r\ll\min\bigl[1,(k/k_0)T^{1/3}\bigr]$, 
we have
\begin{eqnarray}
 G^*(k_i)\,G(k_j)+G^*(k_j)\,G(k_i) & \approx & 
2\left[
\left(1-\frac{2}{5}\frac{k_i^2}{k_0^2}T\right)
\left(1-\frac{2}{5}\frac{k_j^2}{k_0^2}T\right)
+\frac{1}{9}\left(\frac{k_ik_j}{k_0^2}\right)^3T^2
\right]\, ,\nonumber\\
 H(k_k) & \approx & 
  \frac{1}{5}\frac{1}{k_0^2}T\, .
  \label{eqn:GHapprox}
\end{eqnarray}

The fact that the power spectrum (\ref{eqn:amplifiedPR}) is
in excellent agreement with the exact result~\cite{Leach:2001zf}
suggests that the above result (\ref{eqn:GHapprox}) based on the same 
approximation should also give a good estimate even for $Tk^2/k_0^2\gg 1$
as long as $\eta_k<\eta_0$. 
Therefore $B_{\zeta}$ and $f_{NL}$ may be evaluated as
\begin{eqnarray}
B_{\zeta}(k_1,k_2,k_3) & \approx & 
\frac{8\pi^4({\cal P}^{(0)}_{{\cal R}})^2T}{5k_1^3k_2^3k_3^3}
\left\{
\left[
\left(1-\frac{2}{5}\frac{k_1^2}{k_0^2}T\right)
\left(1-\frac{2}{5}\frac{k_2^2}{k_0^2}T\right)
+\frac{1}{9}\left(\frac{k_1k_2}{k_0^2}\right)^3T^2
\right]
\right.
\nonumber\\
 & & 
  \left.
  \times
\left[5(k_1^2+k_2^2)-k_3^2\right]\frac{k_3^3}{k_0^2}
+ \mbox{cyclic}
\right\}\, ,\nonumber\\
f_{NL}(k_1,k_2,k_3)
& \approx& 
\frac{T}{6}
\left\{
\left[
\left(1-\frac{2}{5}\frac{k_1^2}{k_0^2}T\right)^2
+\frac{1}{9}\frac{k_1^6}{k_0^6}T^2
\right]
\left[
\left(1-\frac{2}{5}\frac{k_2^2}{k_0^2}T\right)^2
+\frac{1}{9}\frac{k_2^6}{k_0^6}T^2
\right]
k_3^3
+ \mbox{cyclic}
\right\}^{-1}
\nonumber\\
& & \times
\left\{
\left[
\left(1-\frac{2}{5}\frac{k_1^2}{k_0^2}T\right)
\left(1-\frac{2}{5}\frac{k_2^2}{k_0^2}T\right)
+\frac{1}{9}\left(\frac{k_1k_2}{k_0^2}\right)^3T^2
\right]
\right.
\nonumber\\
 & & 
  \left.
  \times
\left[5(k_1^2+k_2^2)-k_3^2\right]
\frac{k_3^3}{k_0^2}
+ \mbox{cyclic}
\right\}\, .
\label{eq:fNL}
\end{eqnarray}
We expect the above expressions to be fairly good approximations
for the entire wavenumbers in the range $k/k_0<1$ and for any
value of $T$ as long as it is large, 
though it cannot be justified in the rigorous sense.

Setting $k=k_1=k_2=k_3$, we find
\begin{equation}
f_{NL}^{eql}(k) \approx
\frac{3}{2}\frac{k^2}{k_0^2}T
\left[
\left(1-\frac{2}{5}\frac{k^2}{k_0^2}T\right)^2
+ \frac{1}{9}\frac{k^6}{k_0^6}T^2
\right]^{-1}\,,
\label{eq:fNLeql}
\end{equation}
where $f_{NL}^{eql}(k)$ denotes $f_{NL}$ for an equilateral triangle.
For a fixed $T$, it takes a maximum value 
$(f_{NL}^{eql})_{\rm max}=54T/25$ at $k^2/k_0^2=5/(2T)$.
Fig.~1 shows $f_{NL}^{eql}$ as a function of $y=\sqrt{T}k/k_0$ for
$T=10^2$. Note that $f_{NL}^{eql}(k)$ is always positive in this limit.

For comparison, let us also take the limit of a squeezed triangle,
$k=k_1=k_2$, $k_3=0$. In this limit, $f_{NL}$ becomes
\begin{equation}
f_{NL}^{sqz}(k) \approx
\frac{2}{3}\frac{k^2}{k_0^2}T
\left[
\left(1-\frac{2}{5}\frac{k^2}{k_0^2}T\right)^2
+\frac{1}{9}\frac{k^6}{k_0^6}T^2
\right]^{-1}
\left(1-\frac{2}{5}\frac{k^2}{k_0^2}T
 + \frac{1}{9}\frac{k^6}{k_0^6}T^2\right)\, ,
\label{eq:fNLsqz}
\end{equation}
where $f_{NL}^{sqz}(k)$ denotes $f_{NL}$ for a squeezed triangle.
Fig.~2 shows $f_{NL}^{sqzl}$ as a function of $y=\sqrt{T}k/k_0$ for
$T=10^2$. Note that $f_{NL}^{sqz}(k)$ becomes negative for large $T$. 
%%%%%%%%%%%%%%%%%%%%%%%%%%%%%%%%%%%
\begin{figure}[thb]
\begin{center}
\includegraphics[width=10cm]{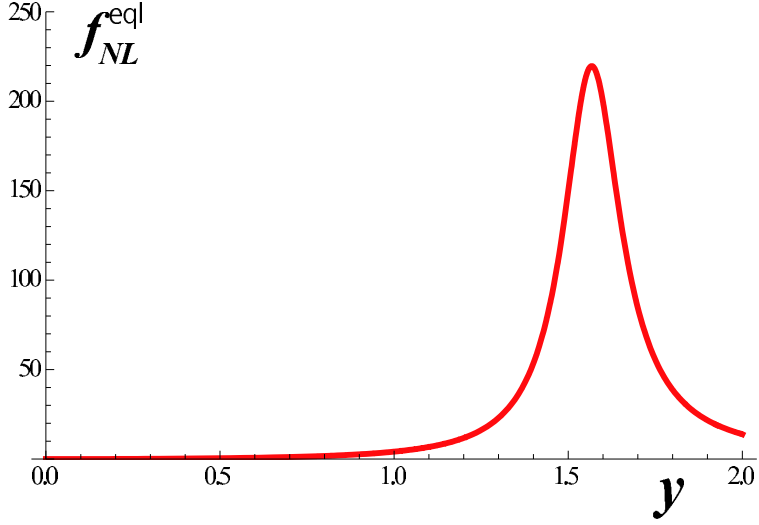}
\end{center}
\caption{$f^{eql}_{NL}(k)$ as a function of $y=\sqrt{T} k/k_0$ for $T=10^2$. }
\end{figure}

\begin{figure}[thb]
\begin{center}
\includegraphics[width=10cm]{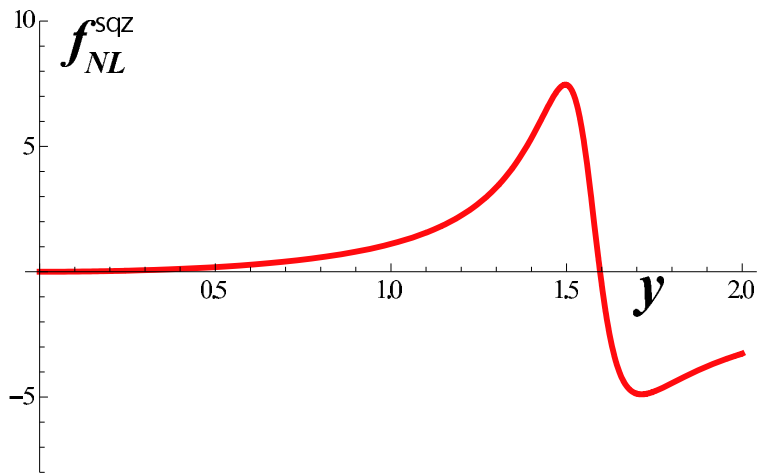} 
\end{center}
\caption{The same as Fig.~1 but for $f^{sqz}_{NL}(k)$.}
\end{figure}

\begin{figure}[thb]
\begin{center}
\begin{tabular}{ll}
\hspace{-0.3cm}
\includegraphics[width=9cm]{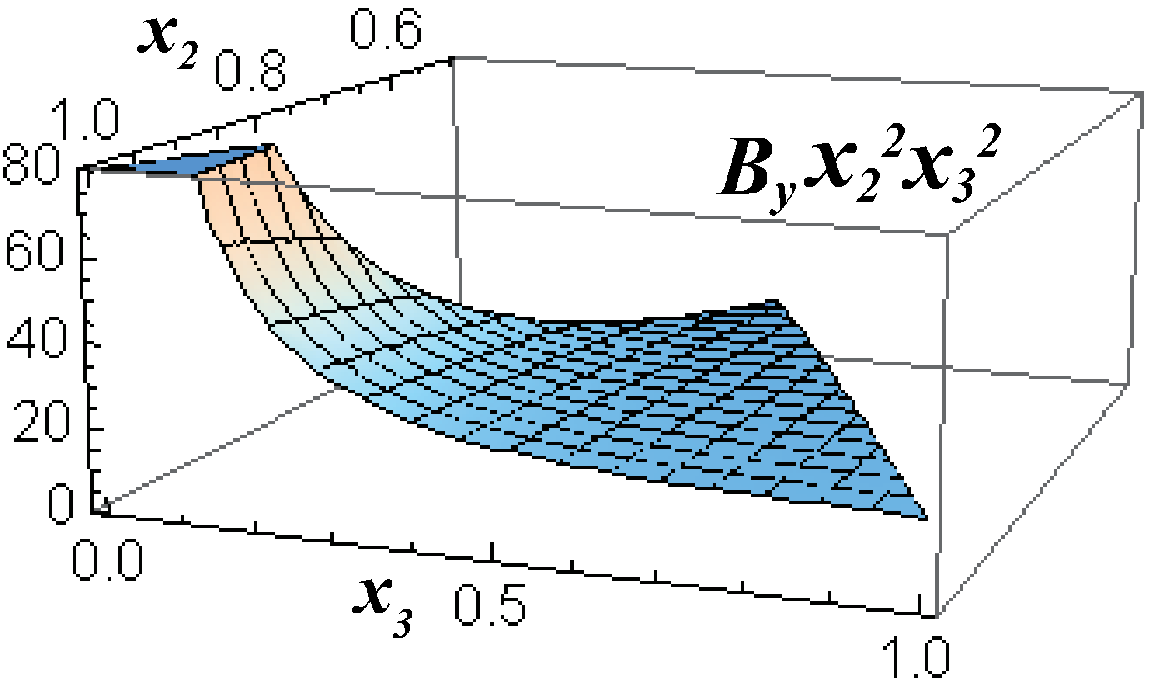}  &
\hspace{-1.cm}
\includegraphics[width=9cm]{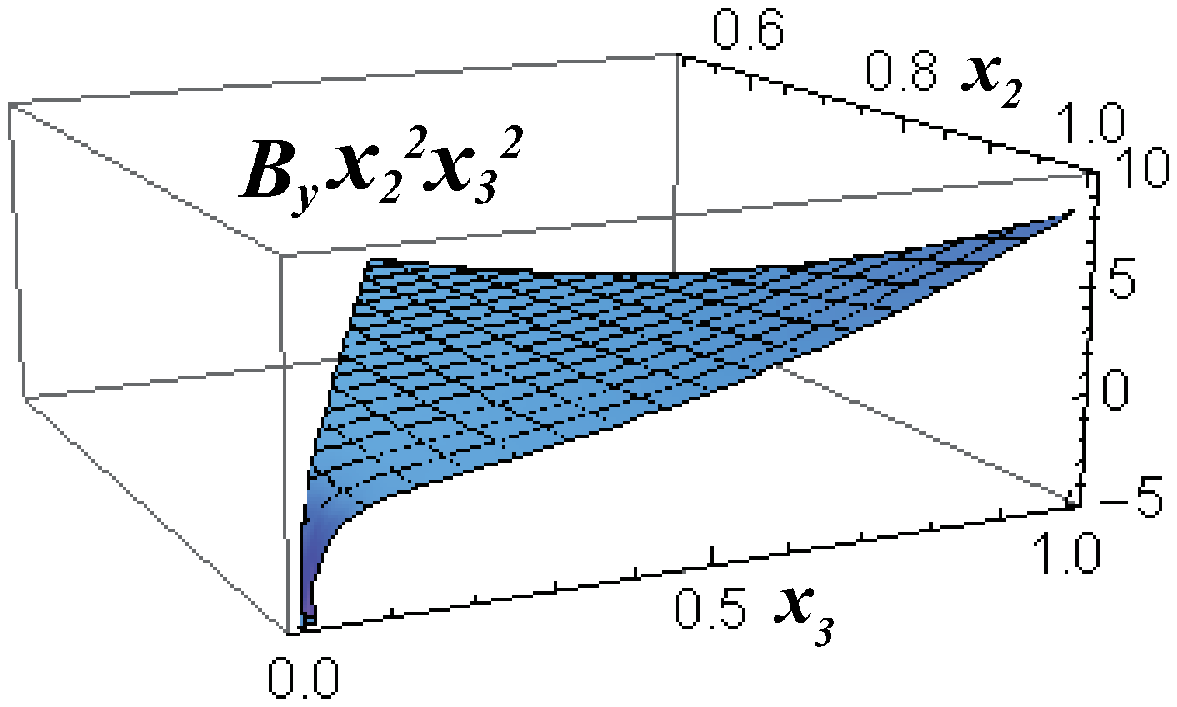}
\end{tabular}
\vspace{-2cm}
\end{center}
\caption{${\cal B}_{y}(1,x_2,x_3)x_2^2 x_3^2$
 as a function of $x_2$ and $x_3$ for $y=1$ (left) and $y=10$ (right).
The bispectrum has a peak at the squeezed shape ($x_2=1$, $x_3=0$)
for $y=1$, while it has a peak at the equilateral shape ($x_2=x_3=1$) 
and a negative peak at the squeezed shape for $y=10$.}
\end{figure}
%%%%%%%%%%%%%%%%%%%%%%%%%%%%%%%%%%%%%%%%%%%%%%%%%

To see the shape dependence of $B_\zeta$, it is convenient to 
define a dimensionless function,
\begin{eqnarray}
 {\cal B}_y(x_1,x_2,x_3)  & \equiv &
  \frac{5k_0^6}
  {8\pi^4({\cal P}^{(0)}_{{\cal R}})^2T^3}
  B_{\zeta}(k_0yx_1/\sqrt{T},k_0yx_2/\sqrt{T},k_0yx_3/\sqrt{T})
\nonumber\\
& \approx &
\frac{1}{y^4x_1^3x_2^3x_3^3}
 \left[
\left(1-\frac{2}{5}y^2x_1^2\right)
\left(1-\frac{2}{5}y^2x_2^2\right)
\left[5(x_1^2+x_2^2)-x_3^2\right]x_3^3
+ \mbox{cyclic}
\right]\, ,
\end{eqnarray}
and identify the dimensionless variables as
\begin{equation}
 y = \sqrt{T}k_1/k_0\,,\quad
x_1=1\,,\quad
x_2=k_2/k_1\,,\quad
x_3=k_3/k_1\,,
\end{equation}
so that
\begin{equation}
 B_{\zeta}(k_1,k_2,k_3) = 
  \frac{8\pi^4({\cal P}^{(0)}_{{\cal R}})^2T^3}{5k_0^6}
  {\cal B}_y(1,x_2,x_3).
\end{equation}
Without loss of generality, we
can restrict our attention to the region $1-x_2<x_3<x_2<1$. 
We plot ${\cal B}_y(1,x_2,x_3)x_2^2 x_3^2$ for $y=1$ and
$y=10$ as a function of $x_2$ and $x_3$ in Fig.~3. For small $y$, we see
that the bispectrum has a peak at the squeezed shape. On the other hand,
for larger $y$, there is a positive peak at the equilateral shape as
well as a negative peak at the squeezed shape.

%%%%%%%%%%%%%%%%%%%%%%%%%%%%%%%%%%%%%%%%%%%%%%%%%
\section{Summary and discussion}
\label{sec:summary}

We have developed a theory of nonlinear cosmological perturbations on
 superhorizon scales for a single scalar field 
 with a general kinetic term and a general form of the potential
to second-order in the spatial gradient expansion. 
The solution to this order is necessary to evaluate correctly
the final amplitude of the curvature perturbation for models of 
inflation with a temporary violation of the slow-roll condition.
 We have employed the ADM formalism and obtained 
the general solution for full nonlinear curvature 
 perturbations valid up through second-order in the gradient expansion.
We have introduced a reasonable variable that represents 
 the nonlinear curvature perturbation on comoving slices ${\cal R}_
c^{\rm NL}$, 
 which reduces to the comoving curvature perturbation ${\cal R}_c^{\rm Lin}$
in the linear limit. Then we have found that ${\cal R}_c^{\rm NL}$
 satisfies a nonlinear second-order differential equation,
 (\ref{eq: basic eq for NL}), as a natural extension of the linear
second-order differential equation.

Then we have formulated the matching of the nonlinear solution
to a perturbative solution at $n$-th-order, ${\cal R}_c^{\rm pert}$, 
on superhorizon scales,
and obtained a formula for the final value of the nonlinear curvature 
 perturbation expressed in terms of ${\cal R}_c^{\rm pert}$ and its
time derivative at the time of matching. Since the evolution of
superhorizon curvature perturbations is genuinely due to the $O(\epsilon^2)$
effect, our formulation can be used to calculate the primordial
 non-Gaussianity beyond the $\delta N$ formalism which is
equivalent to leading order in the gradient expansion. 

Then we have considered the case when the linear approximation
is valid up to the time of horizon crossing for wavenumbers of physical
interest. In this case, we have developed a method to determine
quantities corresponding to ${\cal R}_c^{\rm pert}$ and its time derivative
at the matching time in terms of the linear solution. 

As an example of such a case, we have investigated Starobinsky's
model~\cite{Starobinsky:1992ts} in which  
there is a temporary non-slow-roll stage during inflation
due to a sudden change of the potential slope.
We have found that non-Gaussianity can become large if
the parameter $T\approx {A_+/A_-}$, which characterises
the ratio of the slope before and after the transition,
is large. For $T\gg1$, we have found that the non-Gaussianity parameter for
the bispectrum $f_{NL}({ k}_1,{k}_2,{k}_3)$ 
is peaked at the wavenumbers forming an equilateral triangle,
$k=k_1=k_2=k_3$, denoted by $f_{NL}^{eql}(k)$. It is found to be
positive and takes the maximum value $f_{NL}^{eql}(k)\simeq 2 T$ 
at $y=\sqrt{T}k/k_0\simeq 1.5$ where $k_0$ is the
comoving wavenumber that crosses the horizon at the time
when the potential slope changes.
This implies that, even for a relatively small $T$, say for $T=10$,
it is possible to generate a fairly large non-Gaussianity 
$f_{NL}\sim 20$ at wavenumber $k\simeq 0.5 k_0$.

Our formalism can be applied to many other interesting
circumstances in which the slow-roll condition is temporarily
violated. To mention a couple of examples, a non-slow-roll
stage appears in a double inflation 
model~\cite{Saito:2008em} or in a specific case of 
DBI inflation~\cite{Jain:2007au}.
It is of interest to investigate the non-Gaussianity 
in these models by applying our formalism.

It is also of interest to investigate the case when there is a step
in the inflaton potential instead of a change in the slope,
which was proposed to explain the `features' in the cosmic microwave
background anisotropy~\cite{Chen:2006xjb,Joy:2008qd}.
The case of time-varying sound speed for models with
non-canonical kinetic terms~\cite{Khoury:2008wj}
 may also deserve future study, since a rapid temporal 
variation of the sound velocity violates a certain type
of the slow-roll condition.

Finally, here we have focused on the case of a single scalar field.
An immediate issue is to extend the present formalism to the case of
a multi-component scalar field. We plan to work on this and hope to
report the result in the near future.

%%%%%%%%%%%%%%%%%%%%%%%%%%%%%
\acknowledgments
YuT would like to thank Jun'ichi Yokoyama, Alexei Starobinsky, 
Shuichiro Yokoyama, Ryo Saito and Masahiro Nakashima for their comments 
and discussions on this work. YuT also wishes to acknowledge 
financial support by the Research Center of the Early Universe (RESCEU), 
University of Tokyo and by JSPS Grant-in-Aid for Young Scientists
(B) No.~21740192. 

%%%%%%%%%%%%%%%%%%%%%%%%%%%%%
The work of SM is supported by JSPS Grant-in-Aid for Young Scientists (B) 
No.~17740134, JSPS Grant-in-Aid for Creative Scientific Research No.~19GS0219,
MEXT Grant-in-Aid for Scientific Research on Innovative Areas No.~21111006,
JSPS Grant-in-Aid for Scientific Research (C) No.~21540278, the Mitsubishi
Foundation, and World Premier International Research Center Initiative. 

%%%%%%%%%%%%%%%%%%%%%%%%%%%%%
MS and YoT are supported in part by JSPS Grant-in-Aid for 
Scientific Research (A) No.~21244033,
by JSPS Grant-in-Aid for Creative Scientific Research No.~19GS0219,
and by MEXT Grant-in-Aid for the global COE program at Kyoto University,
``The Next Generation of Physics, Spun from Universality and Emergence''.

%%%%%%%%%%%%%%%%%%%%%%%%%%%%%%%%%%%%%%%%%%%%%%%%
\appendix
\section{Useful formulas from background equations}
\label{apx:formulas}

In this Appendix,
we derive some useful formulas that are used in Sec.~\ref{subsec:nleq}
to modify the apparent forms of the functions $f_K(t)$ and $f_C(t)$
defined in (\ref{eq: f-K t}) and (\ref{eq: f-C t}).

Using the background equations,
\begin{eqnarray}
\dot{H}=-{\kappa^2 \over 2}(\rho+P)\,,
\quad
\frac{d}{dt}\left({1\over \rho+P}\right)={3(1+c_s^2)H \over \rho+P}
-{\rho \Gamma \dot{\phi}\over (\rho+P)^2}\,,
\label{eq: useful relation}
\end{eqnarray}
we obtain a useful formula by integrating the time derivative
of $(a/H)$,
\begin{eqnarray}
\int_{t_*}^{t} a(t') dt'=\biggl[{a\over H}\biggr]^{t}_{t_*}-{\kappa^2 \over 2}
\int_{t_*}^{t} {a(\rho+P)\over H^2} dt'\,.
\label{eq: integralby part}
\end{eqnarray}
Then using the quantity $z$ defined in (\ref{def: variable-z}),
\begin{eqnarray}
z={{a\over H}\left(\rho+P \over c_s^2\right)^{1\over 2}}\,,
\end{eqnarray}
the above can be further transformed to
\begin{eqnarray}
{d\over dt}{\left({1\over a(t)}
\int_{t_*}^{t} a(t') dt'\right)}
={\kappa^2 H\over 2 a}\int_{\eta_*}^{\eta} 
z^2 c_s^2(\eta') d\eta' +{a_* H(t) \over a(t) H_* }\,,
\label{eq: integralby part2}
\end{eqnarray}
where the subscript $*$ indicates that the
quantity is estimated at $t=t_*$ (or $\eta=\eta_*$). 
With the help of (\ref{eq: useful relation}) 
and (\ref{eq: integralby part2}), we obtain 
\begin{eqnarray}
\int_{t_*}^{t}{H \rho\Gamma \dot{\phi}\over 2(\rho+P)^2 a^3}\,(t')dt'
\int_{t_*}^{t'}
a(t'') dt''&=&-{H\over 2(\rho+P) a^3 } 
\int_{t_*}^t a(t') dt' 
+
\int_{t_*}^{t}{dt'\over (\rho+P)}\,\partial_{t'}
\left({H(t')\int_{t_*}^{t'} a(t'') dt''\over 2 a^3(t')}\right)
\nonumber\\
&&+
\int_{t_*}^{t} {3(1+c_s^2) H^2\over 2(\rho+P) a^3}(t') dt'
\int_{t_*}^{t'} a(t'')dt'',
\end{eqnarray}
where the second term in the right hand side of the above equation 
can be rewritten as
\begin{eqnarray}
\int_{t_*}^{t}{dt'\over (\rho+P)}\,\partial_{t'}
\left({H\over 2 a^2}\, {\int_{t_*}^{t'} a(t'') dt''
\over a(t')}\right)
&=&-{\kappa^2\over 4}\int_{t_*}^{t}
{dt'\over a^3(t')}\int_{t_*}^{t'} a(t'') dt''
-\int_{t_*}^{t}
{H^2\over (\rho+P)a^3}(t') dt'
\int_{t_*}^{t'} a(t'') dt'' \nonumber\\
&&+{\kappa^2\over 4}
\int_{\eta_*}^{\eta} {d\eta'\over z^2 c_s^2(\eta')}
\int_{\eta_*}^{\eta'} 
z^2 c_s^2 (\eta'') d\eta''  
+\int_{\eta_*}^{\eta} {d\eta'
\over 2 z^2 c_s^2(\eta')}
{a_*\over H_*},
\end{eqnarray}
and similarly, we obtain
\begin{eqnarray}
\int_{t_*}^{t}{H \rho\Gamma \dot{\phi}\over 2(\rho+P)^2 a^3}(t')dt'
=-{H\over2(\rho+P)a^3 }\bigg|^t_{t_*}
-{\kappa^2\over 4}\int_{t_*}^{t}
{dt'\over a^3(t')}+{3\over 2}\int_{t_*}^{t}
{d\eta' \over z^2(\eta')}\,.
\label{eq: trans-z-relation}
\end{eqnarray}

%%%%%%%%%%%%%%%%%%%%%%%%%%%%%%%%%%%%%%%%%%%%%%%%%%
\section{$t_*$-independence in Linear theory}
\label{app:alphau}

In the linear theory the curvature perturbation 
${\cal R}^{\rm Lin}_{c,\bm{k}}(\eta)$ in the Fourier space satisfies 
%============< EQUATION >==============%
%
\begin{equation}
{{\cal R}^{\rm Lin}_{c,\bm{k}}}''
 + 2\frac{z'}{z}{{\cal R}^{\rm Lin}_{c,\bm{k}}}'
 + c_s^2k^2{\cal R}^{\rm Lin}_{c,\bm{k}} = 0, \label{eqn:RLin-eq}
\end{equation}
%======================================%
and is related to $u^{(0)}_{\bm{k}}$ up to $O(k^2)$ as
%============< EQUATION >==============%
%
\begin{eqnarray}
 {\cal R}^{\rm Lin}_{c,\bm{k}}(\eta) & = & 
  \left[ \alpha^{\rm Lin}_{\bm{k}} + (1-\alpha^{\rm Lin}_{\bm{k}})
   \frac{D(\eta)}{D_*}
   - \left( \frac{F_*}{D_*}D(\eta)-F(\eta) \right)k^2
   + O(k^4) \right] u^{(0)}_{\bm{k}} \nonumber\\
 & = & 
  \left[ 1 - \beta^{\rm Lin}_{\bm{k}}
   \left( 1 -  \frac{D(\eta)}{D_*} \right)
   - \left( \frac{F_*}{D_*}D(\eta)-F(\eta) \right)k^2
   + O(k^4) \right] u^{(0)}_{\bm{k}},  \label{eqn:RLin-u0}
 \end{eqnarray}
%======================================%
where
%============< EQUATION >==============%
%
\begin{equation}
 D(\eta) = 3{\cal H}_*\int_{\eta}^0
  \frac{z^2(\eta_*)}{z^2(\eta')}d\eta', \quad
  F(\eta) = \int_{\eta}^0\frac{d\eta'}{z^2(\eta')}
  \int_{\eta_*}^{\eta'}z^2(\eta'')c_s^2(\eta'')d\eta'',
\end{equation}
%======================================%
%============< EQUATION >==============%
%
\begin{equation}
 D_*=D(\eta_*), \quad F_* = F(\eta_*), 
\end{equation}
%======================================%
and 
%============< EQUATION >==============%
%
\begin{equation}
 \beta^{\rm Lin}_{\bm{k}}
  = 1- \alpha^{\rm Lin}_{\bm{k}}
  = -\frac{D_*}{3{\cal H}_*}
  \left.
   \frac{{{\cal R}^{\rm Lin}_{c,\bm{k}}}'}{{\cal R}^{\rm Lin}_{c,\bm{k}}}
   \right|_{\eta=\eta_*}
  + k^2 F_*.
\end{equation}
%======================================%
In our paper we have assumed that
%============< EQUATION >==============%
%
\begin{equation}
 \beta^{\rm Lin}_{\bm{k}} = O(k^2). 
\end{equation}
%======================================%

By inverting the relation (\ref{eqn:RLin-u0}) and setting $\eta=\eta_k$,
$u^{(0)}_{\bm{k}}$ is expressed as 
%============< EQUATION >==============%
%
\begin{equation}
 u^{(0)}_{\bm{k}} = 
  \left[ 1 + \beta^{\rm Lin}_{\bm{k}}
   \left( 1 -  \frac{D_k}{D_*} \right)
   + \left( \frac{F_*}{D_*}D_k-F_k \right)k^2
   + O(k^4) \right] {\cal R}^{\rm Lin}_{c,\bm{k}}(\eta_k), 
\end{equation}
%======================================%
where
%======================================%
%============< EQUATION >==============%
%
\begin{equation}
 D_k=D(\eta_k), \quad F_k = F(\eta_k). 
\end{equation}
%======================================%
Hence, 
%============< EQUATION >==============%
%
\begin{eqnarray}
 \alpha^{\rm Lin}_{\bm{k}}u^{(0)}_{\bm{k}} & = &
  \left[ 1 - \frac{D_k}{D_*} \beta^{\rm Lin}_{\bm{k}}
   + \left( \frac{F_*}{D_*}D_k-F_k \right)k^2
   + O(k^4) \right] {\cal R}^{\rm Lin}_{c,\bm{k}}(\eta_k)
  \nonumber\\
 & = & 
  \left[ 1 + \frac{D_k}{3{\cal H}_*}
  \left.
   \frac{{{\cal R}^{\rm Lin}_{c,\bm{k}}}'}{{\cal R}^{\rm Lin}_{c,\bm{k}}}
   \right|_{\eta=\eta_*}
  - k^2 F_k 
  + O(k^4) \right] {\cal R}^{\rm Lin}_{c,\bm{k}}(\eta_k)
  \nonumber\\
 & = & 
  \left\{ 1 - k^2 \tilde{F}_k 
   + \left[
      k^2\int_{\eta_k}^{\eta_*}z^2(\eta'')c_s^2(\eta'')d\eta''
      + 
      \left.
       \frac{z^2{{\cal R}^{\rm Lin}_{c,\bm{k}}}'}
       {{\cal R}^{\rm Lin}_{c,\bm{k}}}
      \right|_{\eta=\eta_*}\right]
   \int_{\eta_k}^0\frac{d\eta'}{z^2(\eta')}
  + O(k^4) \right\} {\cal R}^{Lin}_{c,\bm{k}}(\eta_k),
\end{eqnarray}
%======================================%
where
%============< EQUATION >==============%
%
\begin{equation}
 \tilde{F}_k = 
  \int_{\eta_k}^0\frac{d\eta'}{z^2(\eta')}
  \int_{\eta_k}^{\eta'}z^2(\eta'')c_s^2(\eta'')d\eta''.
  \label{eqn:tildeFk}
\end{equation}
%======================================%

By using (\ref{eqn:RLin-eq}), it is easy to show that
%============< EQUATION >==============%
%
\begin{equation}
 \left(
  \frac{z^2{{\cal R}^{\rm Lin}_{c,\bm{k}}}'}
  {{\cal R}^{\rm Lin}_{c,\bm{k}}}
 \right)'
 = -k^2z^2c_s^2
 - \left(\frac{z{{\cal R}^{\rm Lin}_{c,\bm{k}}}'}
  {{\cal R}^{\rm Lin}_{c,\bm{k}}}\right)^2
 = -k^2z^2c_s^2 + O(k^4). 
\end{equation}
%======================================%
This implies that
%============< EQUATION >==============%
%
\begin{equation}
 k^2\int_{\eta_k}^{\eta_*}z^2(\eta'')c_s^2(\eta'')d\eta''
  =  
  -     \left.
	 \frac{z^2{{\cal R}^{\rm Lin}_{c,\bm{k}}}'}
	 {{\cal R}^{\rm Lin}_{c,\bm{k}}}
	\right|_{\eta=\eta_*}
  +     \left.
	 \frac{z^2{{\cal R}^{Lin}_{c,\bm{k}}}'}
	 {{\cal R}^{\rm Lin}_{c,\bm{k}}}
	\right|_{\eta=\eta_k}
  + O(k^4).
\end{equation}
%======================================%
Therefore, we obtain
%============< EQUATION >==============%
%
\begin{equation}
 \alpha^{\rm Lin}_{\bm{k}}u^{(0)}_{\bm{k}} = 
  \tilde{\alpha}^{\rm Lin}_{\bm{k}}
  {\cal R}^{\rm Lin}_{c,\bm{k}}(\eta_k)
  + O(k^4),
  %\label{eqn:alphau}
\end{equation}
%======================================%
where
%============< EQUATION >==============%
%
\begin{equation}
 \tilde{\alpha}^{\rm Lin}_{\bm{k}}
  =
  1 - k^2 \tilde{F}_k 
  + 
  \int_{\eta_k}^0\frac{d\eta'}{z^2(\eta')}
  \times
  \left.
   \frac{z^2{{\cal R}^{\rm Lin}_{c,\bm{k}}}'}
   {{\cal R}^{\rm Lin}_{c,\bm{k}}}
  \right|_{\eta=\eta_k},
\end{equation}
%======================================%
and $\tilde{F}_k$ is given by (\ref{eqn:tildeFk}). This explicitly shows
that $\alpha^{\rm Lin}_{\bm{k}}u^{(0)}_{\bm{k}}$ is independent of
$\eta_*$ up through $O(k^2)$.


\begin{thebibliography}{99}
%\cite{Spergel:2006hy}
\bibitem{Spergel:2006hy}
  %\cite{Komatsu:2003fd}
  %\bibitem{Komatsu:2003fd}
  E.~Komatsu {\it et al.}  [WMAP Collaboration],
 %``First Year Wilkinson Microwave Anisotropy Probe (WMAP) Observations: Tests
 % of Gaussianity,''
  Astrophys.\ J.\ Suppl.\  {\bf 148}, 119 (2003)
  [arXiv:astro-ph/0302223].\\
  D.~N.~Spergel {\it et al.}  [WMAP Collaboration],
 %``Wilkinson Microwave Anisotropy Probe (WMAP) three year results:
 % Implications for cosmology,''
  Astrophys.\ J.\ Suppl.\  {\bf 170}, 377 (2007)
  [arXiv:astro-ph/0603449].

%\cite{Komatsu:2010fb}
\bibitem{Komatsu:2010fb}
  E.~Komatsu {\it et al.},
 %``Seven-Year Wilkinson Microwave Anisotropy Probe (WMAP) Observations:
 % Cosmological Interpretation,''
  arXiv:1001.4538 [astro-ph.CO].

%\cite{Bartolo:2004if}
\bibitem{Bartolo:2004if}
  N.~Bartolo, E.~Komatsu, S.~Matarrese and A.~Riotto,
 %``Non-Gaussianity from inflation: Theory and observations,''
  Phys.\ Rept.\  {\bf 402}, 103 (2004)
  [arXiv:astro-ph/0406398].

\bibitem{Komatsu:2001rj}
  E.~Komatsu and D.~N.~Spergel,
 %``Acoustic signatures in the primary microwave background bispectrum,''
  Phys.\ Rev.\  D {\bf 63}, 063002 (2001)
  [arXiv:astro-ph/0005036].

%\cite{:2006uk}
\bibitem{Planck:2006uk}
    [Planck Collaboration],
 %``Planck: The scientific programme,''
  arXiv:astro-ph/0604069.



%\cite{Seery:2005wm}
\bibitem{Seery:2005wm}
  D.~Seery and J.~E.~Lidsey,
 %``Primordial non-gaussianities in single field inflation,''
  JCAP {\bf 0506}, 003 (2005)
  [arXiv:astro-ph/0503692].


%\cite{Rigopoulos:2003ak}
\bibitem{Rigopoulos:2003ak}
  G.~I.~Rigopoulos and E.~P.~S.~Shellard,
 %``The separate universe approach and the evolution of nonlinear  superhorizon
 % cosmological perturbations,''
  Phys.\ Rev.\  D {\bf 68}, 123518 (2003)
  [arXiv:astro-ph/0306620].\\
  G.~I.~Rigopoulos and E.~P.~S.~Shellard,
 %``Non-linear inflationary perturbations,''
  JCAP {\bf 0510}, 006 (2005)
  [arXiv:astro-ph/0405185].

%\cite{Lyth:2005fi}
\bibitem{Lyth:2005fi}
  D.~H.~Lyth and Y.~Rodriguez,
 %``The inflationary prediction for primordial non-gaussianity,''
  Phys.\ Rev.\ Lett.\  {\bf 95}, 121302 (2005)
  [arXiv:astro-ph/0504045]. \\
  %\bibitem{Lyth:2005du}
  D.~H.~Lyth and Y.~Rodriguez,
 %``Non-gaussianity from the second-order cosmological perturbation,''
  Phys.\ Rev.\  D {\bf 71}, 123508 (2005)
  [arXiv:astro-ph/0502578].


%\cite{Seery:2005gb}
\bibitem{Seery:2005gb}
  D.~Seery and J.~E.~Lidsey,
 %``Primordial non-gaussianities from multiple-field inflation,''
  JCAP {\bf 0509}, 011 (2005)
  [arXiv:astro-ph/0506056].


\bibitem{Sasaki:1998ug}
  M.~Sasaki and T.~Tanaka,
 %``Super-horizon scale dynamics of multi-scalar inflation,''
  Prog.\ Theor.\ Phys.\  {\bf 99}, 763 (1998)
  [arXiv:gr-qc/9801017].

%\cite{Lyth:2004gb}
\bibitem{Lyth:2004gb}
  D.~H.~Lyth, K.~A.~Malik and M.~Sasaki,
 %``A general proof of the conservation of the curvature perturbation,''
  JCAP {\bf 0505}, 004 (2005)
  [arXiv:astro-ph/0411220].

%\cite{Langlois:2006vv}
\bibitem{Langlois:2006vv}
  D.~Langlois and F.~Vernizzi,
 %``Nonlinear perturbations of cosmological scalar fields,''
  JCAP {\bf 0702}, 017 (2007)
  [arXiv:astro-ph/0610064].

%\cite{Maldacena:2002vr}
\bibitem{Maldacena:2002vr}
  J.~M.~Maldacena,
 %``Non-Gaussian features of primordial fluctuations in single field
 % inflationary models,''
  JHEP {\bf 0305}, 013 (2003)
  [arXiv:astro-ph/0210603].




%%% 以下コメントアウトするときは  \if0...\fi %%%%%


\if0
%%%%%%%%%% 2nd order perturb %%%%%%%%%%%%%%%%
%\cite{Malik:2003mv}
\bibitem{Malik:2003mv}
  K.~A.~Malik and D.~Wands,
 %``Evolution of second order cosmological perturbations,''
  Class.\ Quant.\ Grav.\  {\bf 21}, L65 (2004)
  [arXiv:astro-ph/0307055].


\bibitem{Acquaviva:2002ud}
  V.~Acquaviva, N.~Bartolo, S.~Matarrese and A.~Riotto,
 %``Second-order cosmological perturbations from inflation,''
  Nucl.\ Phys.\  B {\bf 667}, 119 (2003)
  [arXiv:astro-ph/0209156].  

%\cite{Bartolo:2003gh}
\bibitem{Bartolo:2003gh}
  N.~Bartolo, S.~Matarrese and A.~Riotto,
 %``Enhancement of Non-Gaussianity after Inflation,''
  JHEP {\bf 0404}, 006 (2004)
  [arXiv:astro-ph/0308088].
%%%%%%%%%%%%%%%%%%%%%%%%%%%%%%%%%%%%%%%%%%%%%
\fi


 
  %\cite{Yokoyama:2007uu}
\bibitem{Yokoyama:2007uu}
  S.~Yokoyama, T.~Suyama and T.~Tanaka,
 %``Primordial Non-Gaussianity in Multi-Scalar Slow-Roll Inflation,''
  JCAP {\bf 0707}, 013 (2007)
  [arXiv:0705.3178 [astro-ph]].\\
  %\cite{Yokoyama:2007dw}
%\bibitem{Yokoyama:2007dw}
  S.~Yokoyama, T.~Suyama and T.~Tanaka,
 %``Primordial Non-Gaussianity in Multi-Scalar Inflation,''
  Phys.\ Rev.\  D {\bf 77}, 083511 (2008)
  [arXiv:0711.2920 [astro-ph]].

%\cite{Byrnes:2008wi}
\bibitem{Byrnes:2008wi}
  C.~T.~Byrnes, K.~Y.~Choi and L.~M.~H.~Hall,
 %``Conditions for large non-Gaussianity in two-field slow-roll inflation,''
  JCAP {\bf 0810}, 008 (2008)
  [arXiv:0807.1101 [astro-ph]].\\
%\cite{Byrnes:2009qy}
%\bibitem{Byrnes:2009qy}
  C.~T.~Byrnes and G.~Tasinato,
 %``Non-Gaussianity beyond slow roll in multi-field inflation,''
  JCAP {\bf 0908}, 016 (2009)
  [arXiv:0906.0767 [astro-ph.CO]].

\bibitem{Suyama:2007bg}
  T.~Suyama and M.~Yamaguchi,
 %``Non-Gaussianity in the modulated reheating scenario,''
  Phys.\ Rev.\  D {\bf 77}, 023505 (2008)
  [arXiv:0709.2545 [astro-ph]].

\bibitem{Sasaki:2008uc}
  M.~Sasaki,
 %``Multi-brid inflation and non-Gaussianity,''
  Prog.\ Theor.\ Phys.\  {\bf 120}, 159 (2008)
  [arXiv:0805.0974 [astro-ph]].\\
%\cite{Naruko:2008sq}
%\bibitem{Naruko:2008sq}
  A.~Naruko and M.~Sasaki,
 %``Large non-Gaussianity from multi-brid inflation,''
    Prog.\ Theor.\ Phys.\  {\bf 121}, 193 (2009)
  [arXiv:0807.0180 [astro-ph]]. 

\bibitem{Malik:2006pm} 
  K.~A.~Malik and D.~H.~Lyth,
 %``A numerical study of non-gaussianity in the curvaton scenario,''
  JCAP {\bf 0609}, 008 (2006)
  [arXiv:astro-ph/0604387].

\bibitem{Sasaki:2006kq}
  M.~Sasaki, J.~Valiviita and D.~Wands,
 %``Non-gaussianity of the primordial perturbation in the curvaton model,''
  Phys.\ Rev.\  D {\bf 74}, 103003 (2006)
  [arXiv:astro-ph/0607627]. 


\bibitem{Alishahiha:2004eh}
  M.~Alishahiha, E.~Silverstein and D.~Tong,
 %``DBI in the sky,''
  Phys.\ Rev.\  D {\bf 70}, 123505 (2004)
  [arXiv:hep-th/0404084]. 

\bibitem{Chen:2006nt}
  X.~Chen, M.~x.~Huang, S.~Kachru and G.~Shiu,
 %``Observational signatures and non-Gaussianities of general single field
 % inflation,''
  JCAP {\bf 0701}, 002 (2007)
  [arXiv:hep-th/0605045].

%\cite{Chen:2006xjb}
\bibitem{Chen:2006xjb}
  X.~Chen, R.~Easther and E.~A.~Lim,
 %``Large non-Gaussianities in single field inflation,''
  JCAP {\bf 0706}, 023 (2007) 
  [arXiv:astro-ph/0611645].\\
%\bibitem{Chen:2008wn}
  X.~Chen, R.~Easther and E.~A.~Lim,
 %``Generation and Characterization of Large Non-Gaussianities in Single Field
 % Inflation,''
  JCAP {\bf 0804}, 010 (2008) 
  [arXiv:0801.3295 [astro-ph]].


\bibitem{Wands:2000dp}
  D.~Wands, K.~A.~Malik, D.~H.~Lyth and A.~R.~Liddle,
 %``A new approach to the evolution of cosmological perturbations on large
 % scales,''
  Phys.\ Rev.\  D {\bf 62}, 043527 (2000)
  [arXiv:astro-ph/0003278].

%\cite{Sasaki:1995aw}
\bibitem{Sasaki:1995aw}
  M.~Sasaki and E.~D.~Stewart,
 %``A General Analytic Formula For The Spectral Index Of The Density
 % Perturbations Produced During Inflation,''
  Prog.\ Theor.\ Phys.\  {\bf 95}, 71 (1996)
  [arXiv:astro-ph/9507001].

%\cite{Starobinsky:1986fxa}
\bibitem{Starobinsky:1986fxa}
  A.~A.~Starobinsky,
  %``Multicomponent de Sitter (Inflationary) Stages and the Generation of
  %Perturbations,''
  JETP Lett.\  {\bf 42}, 152 (1985)
  [Pisma Zh.\ Eksp.\ Teor.\ Fiz.\  {\bf 42}, 124 (1985)].

%\cite{Senatore:2009gt}
\bibitem{Senatore:2009gt}
  L.~Senatore, K.~M.~Smith and M.~Zaldarriaga,
 %``Non-Gaussianities in Single Field Inflation and their Optimal Limits from
 % the WMAP 5-year Data,''
  JCAP {\bf 1001}, 028 (2010) 
  [arXiv:0905.3746 [astro-ph.CO]].

\if0
%\cite{Weinberg:2008hq}
\bibitem{Weinberg:2008hq}
  S.~Weinberg,
 %``Effective Field Theory for Inflation,''
  Phys.\ Rev.\  D {\bf 77}, 123541 (2008)
  [arXiv:0804.4291 [hep-th]].
\fi

\if0
%\cite{ArmendarizPicon:1999rj}
\bibitem{ArmendarizPicon:1999rj}
  C.~Armendariz-Picon, T.~Damour and V.~F.~Mukhanov,
 %``k-inflation,''
  Phys.\ Lett.\  B {\bf 458}, 209 (1999)
  [arXiv:hep-th/9904075].


\bibitem{ArkaniHamed:2003uz}
  N.~Arkani-Hamed, P.~Creminelli, S.~Mukohyama and M.~Zaldarriaga,
 %``Ghost inflation,''
  JCAP {\bf 0404}, 001 (2004)
  [arXiv:hep-th/0312100].


 \bibitem{Silverstein:2003hf}
  E.~Silverstein and D.~Tong,
 %``Scalar speed limits and cosmology: Acceleration from D-cceleration,''
  Phys.\ Rev.\  D {\bf 70}, 103505 (2004)
  [arXiv:hep-th/0310221].
\fi


%%%%%%%%%%%%%%%%%%%%%%%%%%%%%%%%%%%%%

%\cite{Nambu:1994hu}
\bibitem{Nambu:1994hu}
  Y.~Nambu and A.~Taruya,
  %``Application of gradient expansion to inflationary universe,''
  Class.\ Quant.\ Grav.\  {\bf 13}, 705 (1996)
  [arXiv:astro-ph/9411013].
%\cite{Taruya:1997iv} ----delta N ?
%\bibitem{Taruya:1997iv}
 % A.~Taruya and Y.~Nambu,
  %``Cosmological perturbation with two scalar fields in reheating after
  %inflation,''
  %Phys.\ Lett.\  B {\bf 428}, 37 (1998)
  %[arXiv:gr-qc/9709035].

\bibitem{Salopek:1990jq}
  D.~S.~Salopek and J.~R.~Bond,
 %``Nonlinear evolution of long wavelength metric fluctuations in inflationary
 % models,''
  Phys.\ Rev.\  D {\bf 42}, 3936 (1990).

%\cite{Kodama:1997qw}
\bibitem{Kodama:1997qw}
  H.~Kodama and T.~Hamazaki,
  %``Evolution of cosmological perturbations in the long wavelength limit,''
  Phys.\ Rev.\  D {\bf 57}, 7177 (1998)
  [arXiv:gr-qc/9712045].



\bibitem{Leach:2001zf}
  S.~M.~Leach, M.~Sasaki, D.~Wands and A.~R.~Liddle,
 %``Enhancement of superhorizon scale inflationary curvature perturbations,''
  Phys.\ Rev.\  D {\bf 64}, 023512 (2001)
  [arXiv:astro-ph/0101406].

%\cite{Seto:1999jc}
\bibitem{Seto:1999jc}
  O.~Seto, J.~Yokoyama and H.~Kodama,
 %``What happens when the inflaton stops during inflation,''
  Phys.\ Rev.\  D {\bf 61}, 103504 (2000)
  [arXiv:astro-ph/9911119].

\bibitem{Jain:2007au}
  R.~K.~Jain, P.~Chingangbam and L.~Sriramkumar,
 %``Amplification of tachyonic perturbations at super-Hubble scales,''
  JCAP {\bf 0710}, 003 (2007)
  [arXiv:astro-ph/0703762].\\
%\cite{Jain:2009pm}
%\bibitem{Jain:2009pm}
  R.~K.~Jain, P.~Chingangbam, L.~Sriramkumar and T.~Souradeep,
 %``The tensor-to-scalar ratio in punctuated inflation,''
  arXiv:0904.2518 [astro-ph.CO].



%\cite{Tanaka:2006zp}
\bibitem{Tanaka:2006zp}
  Y.~Tanaka and M.~Sasaki,
 %``Gradient expansion approach to nonlinear superhorizon perturbations,''
  Prog.\ Theor.\ Phys.\  {\bf 117}, 633 (2007)
  [arXiv:gr-qc/0612191].
  %%CITATION = PTPKA,117,633;%%

%\cite{Tanaka:2007gh}
\bibitem{Tanaka:2007gh}
  Y.~Tanaka and M.~Sasaki,
 %``Gradient expansion approach to nonlinear superhorizon perturbations II   --
 % a single scalar field --,''
  Prog.\ Theor.\ Phys.\  {\bf 118}, 455 (2007)
  [arXiv:0706.0678 [gr-qc]].  

%\cite{Takamizu:2008ra}
\bibitem{Takamizu:2008ra}
  Y.~Takamizu and S.~Mukohyama,
 %``Nonlinear superhorizon perturbations of non-canonical scalar field,''
  JCAP {\bf 0901}, 013 (2009) 
  [arXiv:0810.0746 [gr-qc]].


%\cite{Garriga:1999vw}
\bibitem{Garriga:1999vw}
  J.~Garriga and V.~F.~Mukhanov,
 %``Perturbations in k-inflation,''
  Phys.\ Lett.\  B {\bf 458}, 219 (1999)
  [arXiv:hep-th/9904176].


%\cite{Hamazaki:2008mh}
\bibitem{Hamazaki:2008mh}
  T.~Hamazaki,
 %``Long wavelength limit of evolution of nonlinear cosmological
 % perturbations,''
  Phys.\ Rev.\  D {\bf 78}, 103513 (2008)
  [arXiv:0811.2366 [astro-ph]].

%\cite{Kodama:1985bj}   
\bibitem{Kodama:1985bj}
  H.~Kodama and M.~Sasaki,
 %``Cosmological Perturbation Theory,''
  Prog.\ Theor.\ Phys.\ Suppl.\  {\bf 78} (1984) 1.

%\cite{Mukhanov:1990me}
\bibitem{Mukhanov:1990me}
  V.~F.~Mukhanov, H.~A.~Feldman and R.~H.~Brandenberger,
 %``Theory of cosmological perturbations. Part 1. Classical perturbations. Part
 % 2. Quantum theory of perturbations. Part 3. Extensions,''
  Phys.\ Rept.\  {\bf 215}, 203 (1992).


%\cite{Leach:2000yw}
\bibitem{Leach:2000yw}
  S.~M.~Leach and A.~R.~Liddle,
 %``Inflationary perturbations near horizon crossing,''
  Phys.\ Rev.\  D {\bf 63}, 043508 (2001)
  [arXiv:astro-ph/0010082].

%\cite{Saito:2008em}
\bibitem{Saito:2008em}
  R.~Saito, J.~Yokoyama and R.~Nagata,
 %``Single-field inflation, anomalous 
 %  enhancement of superhorizon fluctuations 
 % and non-Gaussianity in primordial black hole formation,''
  JCAP {\bf 0806}, 024 (2008)
  [arXiv:0804.3470 [astro-ph]].

%\cite{Saito:2008jc}
%\bibitem{Saito:2008jc}
%  R.~Saito and J.~Yokoyama,
% %``Gravitational wave background as a probe of the primordial black hole
%  abundance,''
 % Phys.\ Rev.\ Lett.\  {\bf 102}, 161101 (2009)
%  [arXiv:0812.4339 [astro-ph]].


  
%\cite{Starobinsky:1992ts}
\bibitem{Starobinsky:1992ts}
  A.~A.~Starobinsky,
 %``Spectrum Of Adiabatic Perturbations In The Universe When There Are
 % Singularities In The Inflation Potential,''
  JETP Lett.\  {\bf 55} (1992) 489
  [Pisma Zh.\ Eksp.\ Teor.\ Fiz.\  {\bf 55} (1992) 477].



%\cite{Joy:2008qd}
\bibitem{Joy:2008qd}
  M.~Joy, A.~Shafieloo, V.~Sahni and A.~A.~Starobinsky,
 %``Is a step in the primordial spectral index favored by CMB data?,''
  JCAP {\bf 0906}, 028 (2009)
  [arXiv:0807.3334 [astro-ph]].

%\cite{Khoury:2008wj}
\bibitem{Khoury:2008wj}
  J.~Khoury and F.~Piazza,
 %``Rapidly-Varying Speed of Sound, Scale Invariance and Non-Gaussian
 % Signatures,''
  JCAP {\bf 0907}, 026 (2009)
  [arXiv:0811.3633 [hep-th]].


%\bibitem{Nakashima2010}
%  M.~Nakashima, R.~Saito, Y.~Takamizu and J~Yokoyama, 
%  ``The effect of varying sound velocity 
%on primordial curvature perturbations,'' in preparation.


\end{thebibliography}
\end{document}